\documentclass[sigconf,nonacm]{acmart}
\AtBeginDocument{%
  }

\copyrightyear{2026}
\acmYear{2026}
\acmDOI{XXXXXXX.XXXXXXX}

\acmConference[CHI '26]{Proceedings of the ACM CHI Conference on Human Factors in Computing Systems}{2026}{Vienna, Austria}
\acmISBN{978-1-4503-XXXX-X/26/04}
\usepackage{array}
\usepackage[table]{xcolor}
\usepackage[inkscapelatex=false]{svg}
\usepackage{booktabs}
\usepackage{subcaption}
\usepackage{xspace}
\usepackage{csvsimple}   
\usepackage{totpages}    
\graphicspath{{logs_paper/}{figures/}}
\svgpath{{figures/}} 
\newcommand{\systemname}{\textsc{SimClinician}\xspace}

\makeatletter
\newcommand{\safeinputtab}[1]{%
  \begingroup
  \catcode`\%=12\relax 
  \IfFileExists{#1}{\input{#1}}{\textit{(missing file: #1)}}%
  \endgroup
}
\newcommand{\inputsummary}{%
  \IfFileExists{logs_paper/summary_all.tex}{%
    \input{logs_paper/summary_all.tex}%
  }{%
    \input{summary_all.tex}%
  }%
}
\makeatother

\usepackage{xcolor,graphicx,adjustbox}
\setkeys{Gin}{width=0.6\linewidth}          
\svgsetup{inkscapelatex=false,width=0.6\linewidth} 
\newcommand{\figwidth}{0.6\linewidth}        

\AtBeginEnvironment{table}{\small}  
\usepackage{tikz}
\usetikzlibrary{
  arrows.meta,positioning,fit,shapes,shapes.multipart,shapes.geometric,calc,backgrounds
}

\definecolor{Ink}{HTML}{1C1C1C}
\definecolor{Slate}{HTML}{4C4F69}
\definecolor{Panel}{HTML}{F7F7FA}
\definecolor{Line}{HTML}{D7D8E0}
\definecolor{cText}{HTML}{2E68AD}   
\definecolor{cAudio}{HTML}{17A15B}  
\definecolor{cVideo}{HTML}{C8524D}  
\definecolor{cFuse}{HTML}{7C5DC4}   
\definecolor{cTrust}{HTML}{FFA31A}  
\definecolor{cClin}{HTML}{5B5B5B}   

\tikzset{
  panel/.style = {rounded corners=2mm, draw=Line, fill=Panel,
                  line width=0.6pt, inner sep=6pt, minimum height=6.8cm},
  step/.style  = {rectangle, rounded corners=2mm, draw=#1!70, fill=#1!10,
                  line width=0.7pt, inner sep=3pt, align=center, minimum height=7mm},
  box/.style   = {rectangle, rounded corners=2mm, draw=Ink!75, fill=white,
                  line width=0.6pt, inner sep=3pt, align=center, minimum height=7mm},
  arrow/.style     = {-{Stealth[length=2.8mm,width=2mm]}, line width=0.9pt, draw=Ink!90},
  thinarrow/.style = {-{Stealth[length=2.3mm,width=1.7mm]}, line width=0.7pt, draw=Ink!70},
  tag/.style       = {font=\bfseries\footnotesize, text=Slate},
  tiny/.style      = {font=\scriptsize, text=Slate},
  chip/.style      = {rounded corners=1.6mm, draw=#1!70, fill=#1!10, inner sep=1.2pt}
}
\tikzset{every picture/.style={scale=0.85, every node/.style={transform shape}}}

\usepackage{algorithm}       
\usepackage{algpseudocode}   

\newif\ifalgonumbers
\algonumberstrue             

\ifalgonumbers
  \algrenewcommand\alglinenumber[1]{\footnotesize\sf #1} 
\else
  \algrenewcommand\alglinenumber[1]{}                    
\fi

\algrenewcommand\algorithmicindent{1.2em}
\AtBeginEnvironment{algorithmic}{\small}

\begin{document}

\title[SimClinician]{SimClinician: A Multimodal Simulation Testbed for Reliable Psychologist–AI Collaboration in Mental Health Diagnosis}

\author{Filippo Cenacchi}
\affiliation{%
  \institution{Macquarie University}
  \department{School of Computing}
  \city{Sydney}
  \country{Australia}}
\email{filippo.cenacchi@mq.edu.au}

\author{Longbing Cao}
\affiliation{%
  \institution{Macquarie University}
  \department{School of Computing}
  \city{Sydney}
  \country{Australia}}
\email{longbing.cao@mq.edu.au}

\author{Deborah Richards}
\affiliation{%
  \institution{Macquarie University}
  \department{School of Computing}
  \city{Sydney}
  \country{Australia}}
\email{deborah.richards@mq.edu.au}

\renewcommand{\shortauthors}{Cenacchi et al.}

\begin{abstract}
AI-based mental health diagnosis is often judged by benchmark accuracy, yet in practice its value depends on how psychologists respond—whether they accept, adjust, or reject AI suggestions. Mental health makes this especially challenging: decisions are continuous and shaped by cues in tone, pauses, word choice, and nonverbal behaviors of patients. Current research rarely examines how AI diagnosis interface design influences these choices, leaving little basis for reliable testing before live studies. We present \systemname{}, an interactive simulation platform, to transform patient data into psychologist–AI collaborative diagnosis. Contributions include: (1) a dashboard integrating audio, text, and gaze–expression patterns; (2) an avatar module rendering de-identified dynamics for analysis; (3) a decision layer that maps AI outputs to multimodal evidence, letting psychologists review AI reasoning, and enter a diagnosis. Tested on the E-DAIC corpus (276 clinical interviews, expanded to 480,000 simulations), \systemname{} shows that a confirmation step raises acceptance by ~23\%, keeping escalations below 9\%, and maintaining smooth interaction flow.
\end{abstract}



\begin{teaserfigure}
  \centering
  \includegraphics[width=\textwidth]{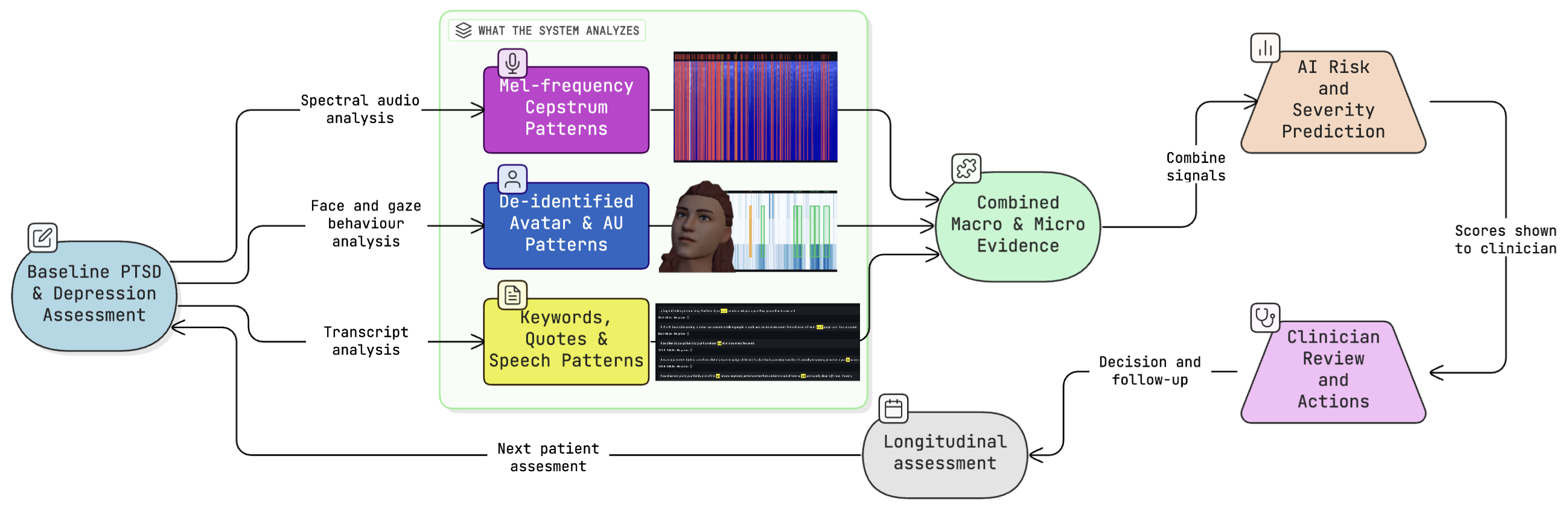}
  \caption{\textbf{System overview.} Baseline assessment feeds audio (MFCC), de-identified avatar (face/gaze), and transcript cues into fused evidence, AI risk/severity prediction, clinician review/actions, and patient longitudinal assessment.}
  \label{fig:system-overview}
\end{teaserfigure}

\maketitle

\section{Introduction}

Artificial intelligence in healthcare is increasingly shifting from model-centric benchmarks to workflow-centric deployments, where predictive outputs are embedded in dashboards, electronic health record (EHR) widgets, and decision-support interfaces reviewed by clinicians. This transition foregrounds a fundamental challenge: clinical AI systems are not judged solely on predictive accuracy but on how effectively their recommendations are negotiated, accepted, or overridden by practitioners operating under time pressure, reputational risk, and ethical responsibility \cite{asan2020artificial,lee2021human}. Mental health triage represents an especially critical domain and poses unique challenges, where diagnostic processes rely on subjective accounts and multimodal behaviors and signals that defy purely numerical decision rules. While public datasets such as DAIC-WOZ and E-DAIC have catalyzed progress in multimodal prediction, most research remains confined to reporting performance metrics, with limited translation into clinical practice \cite{sahili2024multimodal,danner2023advancing}. This gap underscores the pressing need to move beyond benchmark accuracy toward deployment-oriented frameworks.  They must address usability, trust, and real-world integration in high-stakes mental health care and raise a deeper socio-technical question: \emph{what would a clinician actually do with these AI-driven suggestions?} 

Recent research in clinical decision support demonstrates that interface factors such as explanation style, confidence cues, and override frictions profoundly modulate acceptance and deferral behaviors, often more than accuracy alone \cite{bhatt2023collaborative,asan2020artificial,lee2021human}. In mental health diagnosis, where outcomes are rarely binary, the interpretive space widens: a psychologist may confirm an AI’s severity estimate, override upward for patient safety, override downward to avoid false positives, or defer when uncertainty persists. Understanding these dynamics requires methodological vehicles that go beyond accuracy reports. Before recruiting clinicians, an expensive and high-stakes process subject to ethics approvals, researchers need testbeds that can (i) exercise real interfaces end-to-end, (ii) quantify how design choices shift confirm/override patterns, and (iii) produce traceable evidence for institutional review boards. Simulation and methodological testbeds have thus emerged as indispensable buffers between algorithmic development and clinical deployment \cite{ramkumar2024large,weisenburger2024conversational}.

Accordingly, we present \textbf{SimClinician}, a simulation and visualization testbed for studying human–AI collaboration in mental health diagnosis without live participants. SimClinician integrates five components: \textit{an interactive dashboard} that foregrounds multimodal evidence (audio prosody, gaze, transcripts, facial expressions); \textit{a policy-driven simulator} of “synthetic psychologists” that enacts confirm, override, and defer actions; \textit{a headless end-to-end (E2E) validator} that automatically exercises the UI; and \textit{a controller API} that guarantees parity between interactive trials and large-scale batch simulations. In addition, the system introduces an avatar-based visualization of facial and gaze behaviors. Rather than exposing raw video, OpenFace features are mapped onto an animated avatar that reproduces smiles, frowns, blinks, and gaze shifts in real time. This design achieves two goals simultaneously: it makes engagement dynamics perceptible and analyzable in the dashboard, and it protects participant identity by avoiding playback of sensitive recordings. This approach reflects a broader trajectory in HCI and clinical informatics, where privacy-preserving visualization methods have shown that identity can be transformed without sacrificing analytic fidelity. Studies on gaze obfuscation \cite{du2024privategaze}, virtual identity transformation \cite{wang2023identifiable}, and avatar-mediated facial tracking \cite{wei2004avatar}, along with more recent de-identification pipelines in medical contexts \cite{zhu2020deepfakes,david2021privacy}, demonstrate that privacy-by-design can move beyond abstraction to practical, usable systems. By situating our avatar module within this line of work, we treat privacy not as a constraint on analysis but as a design principle that enables responsible multimodal research at scale.

This architecture enables researchers to systematically reproduce judge–advisor dynamics, escalation tendencies, and calibration effects while also surfacing practical issues of latency, selector brittleness, and logging fidelity. In doing so, SimClinician provides both methodological rigor and pragmatic validation that precedes clinician recruitment, offering a principled path toward safer, more trustworthy deployment of clinician-in-the-loop AI.
The methodological novelty of SimClinician lies not in introducing another predictive model, but in elevating the unit of analysis from algorithmic accuracy to \emph{interactional reliability}. By aligning simulated clinician policies with documented phenomena such as confirmation bias, automation bias, and alert fatigue \cite{asan2020artificial,lee2021human}, SimClinician enables researchers to explore how small adjustments to thresholds, explanation granularity, or override frictions propagate into large-scale behavioral outcomes. This shift responds directly to calls from HCI and medical AI research for evaluation frameworks that emphasize use in context rather than AUROC in isolation \cite{bhatt2023collaborative}.

\autoref{fig:system-overview} visualizes the SimClinician architecture. It shows how questionnaire anchors (PHQ-8, PCL-C) are positioned at the entry point, followed by multimodal panels for audio, transcripts, and facial/gaze signals. These converge into a fusion step, which in turn feeds into a measurement-based care loop for longitudinal tracking. This left-to-right choreography captures the intended clinician journey \emph{scan} $\rightarrow$ \emph{probe} $\rightarrow$ \emph{decide} $\rightarrow$ \emph{track} while maintaining API parity between interactive and simulated runs. Presenting this structure early grounds our contributions in a concrete, reproducible workflow. \textit{Note:} \autoref{fig:system-overview} is schematic for legibility—although it shows a single left-to-right journey, the decision layer \emph{branches} into \emph{Accept}, \emph{Override} (up/down), and \emph{Deferral} with distinct logging and follow-up; colors indicate role (green=fusion/evidence, orange=risk prediction, purple=clinician actions, gray=longitudinal tracking), rounded rectangles denote data/report panels rather than decision nodes (diamonds), and the full branching behavior is formalized in Algorithm~\ref{alg:simclinician} and analyzed in Figs.~\ref{fig:policy-mix}--\ref{fig:override-grouped}.
\noindent
While prior work has introduced the notion of ``synthetic clinicians'' in domains such as rehabilitation assessment or treatment acceptance \cite{lee2021human,sivaraman2023ignore}, these efforts typically stopped short of exercising the \emph{actual} interfaces clinicians would use, or relied on limited scenario probes without systematic corpus grounding. \systemname{} differs in three key respects. First, the unified controller guarantees strict UI$\leftrightarrow$API parity: every decision loop that appears in simulation is enacted through the same buttons, selectors, and logs that clinicians would encounter, eliminating the common offline--online drift. Second, the privacy-preserving avatarization module turns raw OpenFace signals into neutral embodied replays, allowing multimodal engagement cues to be analyzed at scale without exposing patient identity. Third, by embedding these mechanisms into a factorial sweep over the E-DAIC corpus, the testbed enables reproducible measurement of policy effects (safety, parsimony, deferral) that would be prohibitively costly in live recruitment.

\textbf{Our contributions are fourfold.} First, we introduce a reproducible simulation methodology for clinician-in-the-loop evaluation using E-DAIC without live human subjects. Second, we design a unified controller that eliminates divergence between interactive experiments and batch-mode simulations, allowing UI studies and high-volume parameter sweeps to inform each other seamlessly. Third, we implement a headless E2E validator that interacts with the real browser dashboard, surfacing interface brittleness early. Finally, we provide a visualization and reporting recipe including judge–advisor statistics, override distributions, calibration diagrams, and simulation logs that empower researchers to quantify the decision impact of interface and policy choices. Together, these contributions establish SimClinician as a rigorous, scalable, and ethically responsible testbed for advancing trustworthy AI-supported mental health diagnosis. \autoref{fig:system-overview} illustrates the overall flow of SimClinician, showing how
baseline questionnaire anchors are connected to multimodal streams and ultimately to
decision and longitudinal measurement based care (MBC) tracking.

\section{Background and Related Work}
\label{sec:background}

Clinical decision support (CDS) has historically been evaluated through predictive performance metrics, yet the true measure of value lies in how recommendations are \emph{interpreted, accepted, overridden, or deferred} by clinicians working under real-world pressures. Recent HCI and medical AI research underscores that biases, interface factors, and socio-cognitive dynamics often outweigh model accuracy in shaping outcomes \cite{asan2020artificial, bhatt2023collaborative, smith2024clinicians}. Building on this insight, we situate \systemname{} within four strands of research: (i) clinician acceptance, override, and deferral in AI-supported decision making; (ii) soft stops and alert fatigue in clinical workflows; (iii) simulation and testbed evaluation of AI systems; and (iv) human–AI biases and interactional effects. These literatures converge on the recognition that trustworthy CDS requires evaluation of the \emph{decision process} itself, not only the statistical properties of algorithms.

\subsection{Clinician Acceptance, Override, and Deferral in AI-Supported Decision Making}
The integration of AI into CDS surfaces critical dynamics of acceptance, override, and deferral that extend far beyond benchmark accuracy. We adopt these three outcomes as the minimal but sufficient set of mutually exclusive actions a clinician can take in response to AI advice, and they directly ground the interface buttons and Algorithm~\ref{alg:simclinician}. Evidence shows that explanation mechanisms can transform interactions from binary compliance into negotiated decision-making, anchoring trust through interpretable outputs and increasing the willingness of clinicians to rely on AI where appropriate \cite{sivaraman2023ignore}. Yet in the absence of normative frameworks, professionals face heightened uncertainty and reputational risk, often leading to resistance against AI recommendations even when technically correct \cite{smith2024clinicians}. Time pressure compounds these dynamics: physicians under cognitive load are more likely to reject algorithmic advice to preserve autonomy, sometimes at the cost of patient outcomes \cite{liang2022save}. Importantly, the framing of AI advice whether delivered as categorical versus probabilistic, or assertive versus conservative, directly influences override rates. Systems optimized for team performance, rather than standalone accuracy, yield stronger alignment and more effective outcomes \cite{bansal2021most}. Notably, assertive or ``zealous'' AI recommendation styles outperform conservative ones in recall-sensitive medical tasks, demonstrating that the calibration of recommendation style itself is a determinant of decision quality \cite{xu2023comparing}. \systemname{} operationalizes these findings by simulating confirmation, override, and deferral policies, allowing us to measure their downstream effects on diagnostic reliability.

\subsection{Soft Stops and Alert Fatigue in Clinical Decision Support}
Within CDS integrated with EHR, the introduction of \textit{soft stops} such as confirmation dialogs, attestation clicks, or lightweight double action mechanisms (brief prompts that require an additional click or attestation before proceeding) has emerged as a middle ground between rigid safety enforcement and workflow efficiency. Unlike hard stops, which disrupt clinical flow and often generate unsafe workarounds, soft stops introduce micro-frictions that nudge users toward safer practices while preserving momentum \cite{sangal2023clinical, fallon2024addressing}. Empirical deployments such as  attestation-based prompts (requiring clinicians to briefly justify an override decision) have been shown to reduce inappropriate prescribing by requiring clinicians to actively justify exceptions \cite{ezran20231190}. Machine-learning–based systems further demonstrate that context-sensitive alerts sustain high acceptance while preventing wrong-drug errors \cite{chen2024ability}. Nevertheless, excessive interruptive alerts remain a persistent liability, with alert fatigue consistently identified as a driver of ignored or unsafe overrides \cite{pourian2025elements, chaparro2022clinical}. To address this, recent designs shift toward passive CDS modalities that lower alert burden while improving outcomes, such as proactive infection control ordering during COVID-19 \cite{fallon2024addressing}. \systemname{} draws on this literature by explicitly modeling confirmation clicks as structured ``soft stops,'' enabling quantitative assessment of how lightweight frictions shape override and acceptance dynamics.

\subsection{Simulation and Testbed Evaluation of Clinical AI}
Because live clinician studies are costly and subject to strict ethical approvals, simulation has become an indispensable stage in the evaluation of high-stakes human–AI infrastructures. Simulated users conceptualized as ``synthetic clinicians'' and dashboard-in-the-loop evaluations provide replicable settings to study how interface design, uncertainty cues, and advice structures influence acceptance and trust calibration before exposing real practitioners. Lee et al.\ demonstrated how interactive AI systems in rehabilitation can anticipate collaborative contingencies through simulation prior to deployment \cite{lee2021human}. Similarly, Sivaraman et al.\ probed clinician acceptance of AI treatment recommendations via simulated decision support, revealing adoption barriers and opportunities for iterative refinement \cite{sivaraman2023ignore}. Beyond interactional logic, technical robustness has also been foregrounded: Ramkumar and Woo advocate for headless UI validation workflows as pre-deployment safeguards against brittle selectors and performance bottlenecks \cite{ramkumar2024large}, while Weisenburger et al.\ show that bot-administered conversational assessments can approximate validated depression interviews, extending scalability while retaining diagnostic validity \cite{weisenburger2024conversational}. Together, these studies highlight simulation as a methodological buffer that both anticipates behavioral dynamics and stress-test interface reliability. Our approach builds on this lineage by unifying headless UI validation, batch-mode policy sweeps, and interactive dashboards within a single controller, ensuring parity between experimental findings and deployment contexts. Such methodological staging has been explicitly endorsed as critical for collaborative AI pipelines \cite{bhatt2023collaborative}, underscoring that trust must be cultivated before, not after, clinical rollout \cite{asan2020artificial}.

\subsection{Clinician-in-the-Loop Decision Support}
Mixed-initiative CDS workflows promise enhanced consistency but necessarily entangle model presentation with human judgment. HCI research argues that interfaces must surface uncertainty, provenance, and rationales, and that evaluation must extend from AUROC to metrics of contextual use \cite{asan2020artificial}. In mental health, where signals are multimodal speech acoustics, linguistic content, gaze direction, and affective displays and decisions are continuous rather than binary, the importance of contextualized presentation is heightened. Rather than discrete labels, clinicians often work with graded severity classes. \systemname{} directly targets this space by simulating continuous depression (0–4) and PTSD (0–2) scales, thereby extending judge–advisor paradigms into domains where subtle gradations, not categorical judgments, determine triage pathways.
Finally, a rich body of HCI and cognitive research documents how interface features and cognitive heuristics shape reliance on AI. Confirmation bias leads humans to preferentially adopt advice that aligns with their priors (existing beliefs or earlier decisions on the same case), even when incorrect, while automation bias fosters over-reliance on machine outputs despite contradictory evidence \cite{asan2020artificial, bansal2021most}. SimClinician translates these theoretical constructs into parameterized policies: confirmation probability, override directionality, and stochastic noise. This design allows biases to be \emph{measured and reproduced}, creating a principled bridge between psychological theory and system-level evaluation.

\paragraph{Summary.} Across the literature, two persistent gaps emerge. First, there is a lack of reproducible testbeds that integrate batch-mode simulation with live UI validation, ensuring that interactional insights are not lost between research and deployment. Second, little attention has been paid to continuous, multimodal decision-making in mental health, where override and deferral dynamics often matter more than AUROC. \systemname{} addresses both by operationalizing insights from HCI, psychology, and medical AI into a reproducible workflow that foregrounds trust, bias, and interactional reliability.

\section{Clinician Workflow and Interface}
\label{sec:workflow}

The heart of \systemname{} lies in the principle that clinical AI must be evaluated not simply as a set of predictive outputs but as a negotiated process between algorithm and human. This negotiation is formally encoded in Algorithm~\ref{alg:simclinician}, which serves as the unifying controller across both the interactive dashboard and batch-mode simulations. Every decision confirmation, override, or deferral is generated through a shared loop that combines model outputs $(\hat d, \hat p)$, derived multimodal evidence, and a parameterized clinician policy $\pi$. The latter specifies thresholds, override priors, and stochastic noise, thereby operationalizing well-documented decision phenomena such as confirmation bias, automation bias, and bounded rationality under time pressure \cite{asan2020artificial,bansal2021most,liang2022save}. By making these dynamics explicit, the algorithm ensures that both experimental results and simulation sweeps reflect the same cognitive levers that govern real-world clinician–AI interaction.

\begin{algorithm}[t]
\caption{Clinician-in-the-loop simulation on E-DAIC (concise)}
\label{alg:simclinician}
\begin{algorithmic}[1]

\Require Cohort $\mathcal{S}$; predictions $P$ with $(\hat d,\hat p)$; metadata $M$; evidence $\{\textsc{AudioSpec},\textsc{TranscriptPhrases},\textsc{GazeScatter}\}$; controller $\mathcal{G}$; policy $\pi=\{\tau_d,\tau_p,b_{\uparrow},b_{\downarrow},\epsilon,\gamma\}$; log $\mathcal{L}$
\Ensure Accepted $(d^\star,p^\star)$, risk $R^\star$, JSONL entry, UI/API parity

\Statex \textbf{Defs:} $\textsc{Risk}(d,p)=100(0.6d/4+0.4p/2)$; 
$\textsc{Clamp}(x,\ell,u)=\min(\max(x,\ell),u)$

\Statex \textbf{ApplyAction} $(d,p,a)$:
\[
(d',p')=\begin{cases}
(\textsc{Clamp}(d{+}1,0,4),\textsc{Clamp}(p{+}1,0,2)) & a=\texttt{up}\\
(\textsc{Clamp}(d{-}1,0,4),\textsc{Clamp}(p{-}1,0,2)) & a=\texttt{down}\\
(d,p) & \text{else}
\end{cases}
\]
\Statex Return $(d',p',a\!\neq\!\texttt{confirm},\textsc{Risk}(d',p'))$

\Statex \textbf{DecideAction} $(d,p,R,\pi)$:
$p_{\text{conf}}=\{0.7,0.5,0.2\}$ if $R{\ge}\max(\tau_d,\tau_p)$, $R{\ge}\min$, else;
$s_{\uparrow}=b_{\uparrow}\mathbb{1}[d\!\in\!\{2,3\}\lor p=1]$;
$s_{\downarrow}=b_{\downarrow}\mathbb{1}[d\!\in\!\{0,1\},p=0]$;
$\tilde s=(s_{\downarrow}{+}\epsilon U_1,p_{\text{conf}{+}\epsilon U_2},s_{\uparrow}{+}\epsilon U_3)$;
$\mathbf{p}=\text{Normalize}(\max(\tilde s,10^{-6}))$; $a{\sim}\text{Cat}(\mathbf{p})$; 
if $a$ override and $\mathbf{p}[a]{<}\gamma$, set $a=\texttt{confirm}$.

\For{$(\texttt{dataset},\texttt{pid})\in\mathcal{S}$}
  \State $(\hat d,\hat p)\gets P$;\; $R\gets\textsc{Risk}(\hat d,\hat p)$;\; $\mathcal{E}\gets$ evidence
  \State \textbf{Dashboard:} render $(\hat d,\hat p,R,\mathcal{E})$, embed JSON
  \State $a\gets\textsc{DecideAction}(\hat d,\hat p,R,\pi)$
  \If{UI} click chosen button \Else POST \texttt{/apply}
  \EndIf
  \State $(d^\star,p^\star,\text{ov},R^\star)\gets\textsc{ApplyAction}(\hat d,\hat p,a)$
  \State Append log $\textsc{Log}(\texttt{dataset},\texttt{pid},\hat d,\hat p,R,a,d^\star,p^\star,R^\star,\text{timestamp})$
\EndFor

\State \textbf{Outputs:} override rate, up/down mix, $\Delta R$, calibration curves, pre/post confusion matrices, replay via $\mathcal{L}$

\end{algorithmic}
\vspace{4pt}
\begingroup\footnotesize
\noindent\textbf{Legend.}\;
$\mathcal{S}$: cohort of cases;\;
$P$: prediction store;\;
$(\hat d,\hat p)$: model outputs (dep.\ $0$–$4$, PTSD $0$–$2$);\;
$M$: metadata (PHQ-8, PCL-C, session info);\;
$\mathcal{E}$: evidence (\textsc{AudioSpec}, \textsc{TranscriptPhrases}, \textsc{GazeScatter});\;
$\mathcal{G}$: controller (UI$\leftrightarrow$API parity);\;
$\pi=\{\tau_d,\tau_p,b_{\uparrow},b_{\downarrow},\epsilon,\gamma\}$: policy (thresholds, priors, noise, friction);\;
$\mathcal{L}$: JSONL log;\;
$a\!\in\!\{\texttt{down},\texttt{confirm},\texttt{up},\texttt{deferral}\}$: action;\;
$(d^\star,p^\star)$: final decision;\;
$R,R^\star$: risk before/after;\;
$\textsc{Clamp}(x,\ell,u)$: bound to $[\ell,u]$;\;
$\mathbb{1}[\cdot]$: indicator;\;
$U_i$: uniform noise;\;
$\text{Normalize}(\cdot)$: to probability simplex;\;
$\text{Cat}(\mathbf{p})$: categorical draw;\;
\texttt{ov}: differs from $(\hat d,\hat p)$.
\par\endgroup
\end{algorithm}

Algorithm~\ref{alg:simclinician} is therefore not an implementation detail but the conceptual backbone of the workflow: it guarantees parity between simulation and real interaction, while making trust-related behaviors measurable. Every panel of the dashboard can be read as a materialization of one element of this loop: the rendering of evidence $\mathcal{E}$, the elicitation of an action $a$, the enforcement of override friction $\gamma$, and the structured logging of outcomes. What follows is a sequential walkthrough of the interface, not in terms of what the clinician ``clicks,'' but in terms of why each panel exists, what reasoning process it supports, and how its design is justified by both psychological practice and HCI research.

\subsection*{Step 0: Anchoring with standardized instruments}
\begin{figure}[t]
  \centering
  \includegraphics[width=1.00\linewidth]{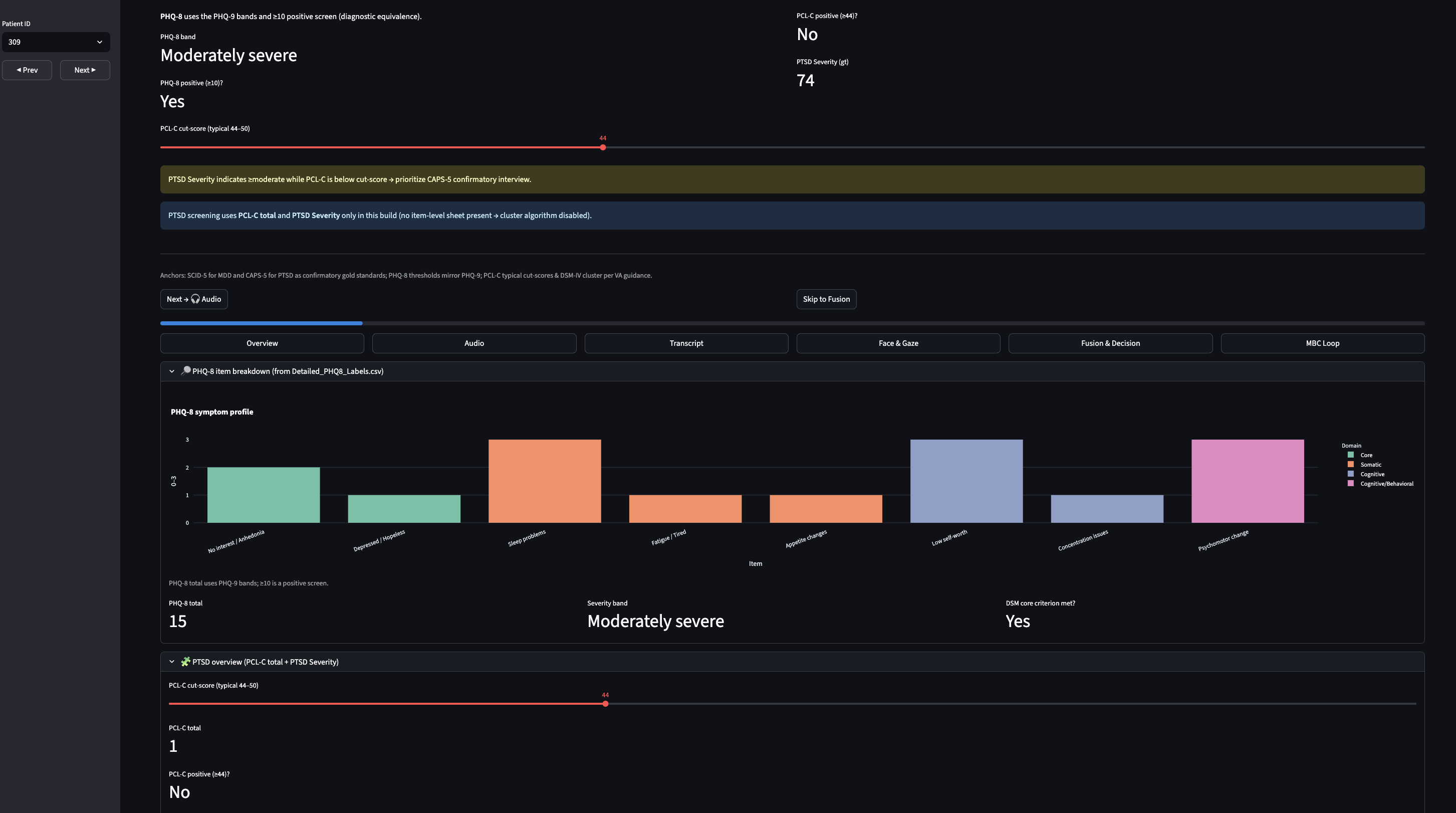}
  \caption{\textbf{Overview panel.} Entry screen showing PHQ-8 totals, PCL-C severity, and cut-off thresholds. These anchors ground subsequent multimodal analysis in validated clinical metrics.}
  \label{fig:overview}
\end{figure}

The journey begins with an overview panel that foregrounds validated self-report scales (PHQ-8, PHQ-9, PCL-C). As shown in Fig.~\ref{fig:overview}, these are not decorative additions but methodological anchors: they embody the principle of MBC, where standardized thresholds structure all downstream reasoning. A PHQ-8 score $\geq 10$ indicates moderate depression \cite{siniscalchi2020depression,trivedi2020can}, while a PCL-C score $\geq 44$ marks probable PTSD \cite{marx2022reliable}. Longitudinal deltas of 5–10 points on the PHQ and 15–20 on the PCL distinguish clinically significant change from measurement noise \cite{blampied2022reliable}. By placing these anchors at the start, the system establishes a shared clinical baseline before multimodal evidence is considered. In our implementation, these cutoffs are stored in $M$ and used to compute \textsc{Risk}$(d,p)$ and to set subsequent policy thresholds.

\subsection*{Step 1: Acoustic markers as perceptual cues}
\begin{figure}[t]
  \centering
  \includegraphics[width=1.00\linewidth]{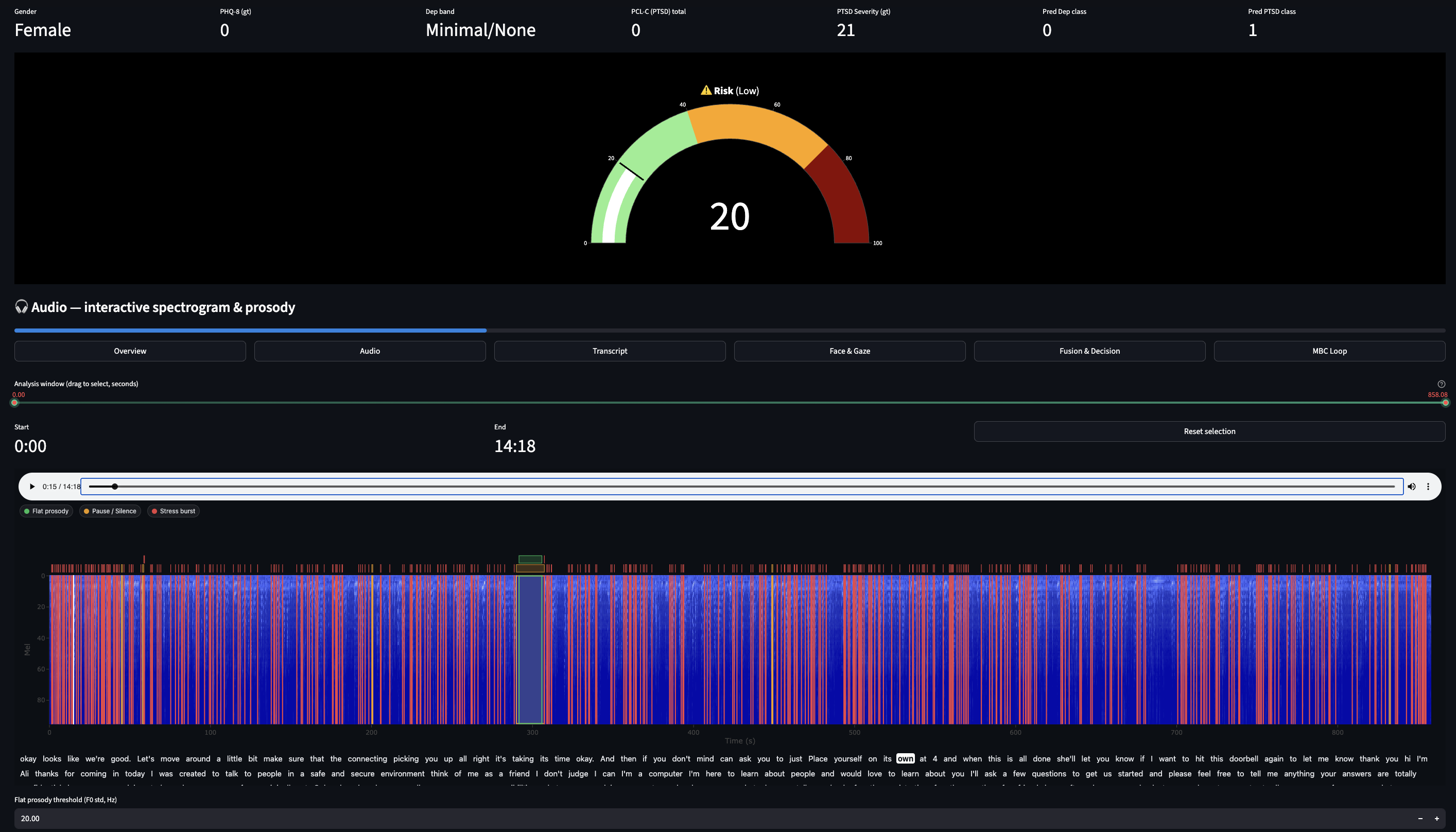}
  \caption{\textbf{Audio panel.} High-resolution spectrogram with rails marking flat prosody, silence, and stress bursts. Threshold sliders encourage parameter probing, making uncertainty explicit.}
  \label{fig:audio}
\end{figure}

From this anchor, the workflow moves into the audio panel (Fig.~\ref{fig:audio}), which surfaces prosodic flatness, prolonged silences, and stress bursts. These are operationalizations of depressive psychomotor slowing and affective reactivity, designed as continuous rails beneath the spectrogram. In algorithmic terms, these acoustic markers form part of the evidence set $\mathcal{E}$, which the controller embeds into the JSON log for each decision. Their visualization reflects principles of ecological interface design: high-contrast rails expose structure in what would otherwise be opaque waveforms. The design choice to make thresholds adjustable reflects the HCI practice of
\emph{parameter-space exploration} (“parameter probing”): exposing tunable
criteria so users can run what-if analyses and test robustness under stricter
settings \cite{sedlmair2014vpsa,wexler2020whatif,marks1997designgalleries,edition2013diagnostic,krause2016predictions}. This responds to empirical evidence that prosody and pauses are among the most reliable nonverbal correlates of major depression, but are often overlooked in purely transcript-based models \cite{trivedi2020implementing,connors2021gets}.

\subsection*{Step 2: Lexical dynamics and temporal focus}
\begin{figure}[t]
  \centering
  \includegraphics[width=1.00\linewidth]{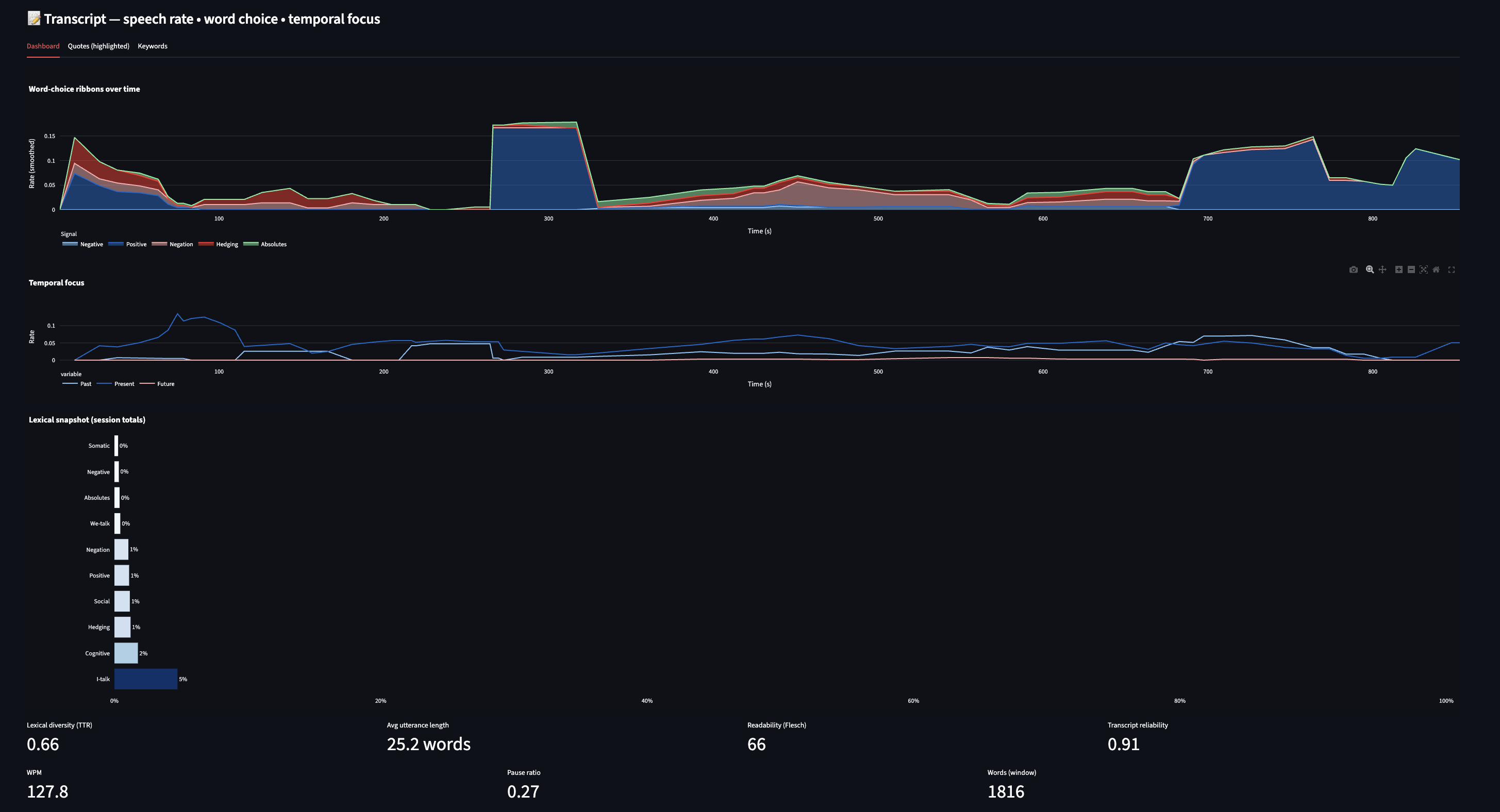}
  \caption{\textbf{Transcript dashboard.} Ribbons show negative/positive language, hedging, and negation over time, alongside a temporal-focus line. Footer metrics summarize lexical diversity, pause ratios, and readability.}
  \label{fig:transcript-dash}
\end{figure}

The transcript dashboard (Fig.~\ref{fig:transcript-dash}) visualizes linguistic \emph{cues} over time—negations, absolutist phrasing, hedging, sentiment polarity, and temporal focus (past/present/future). These are not used as diagnostic criteria; rather, they are theory-aligned markers that prior work has associated with affective state and symptom burden in depression and related conditions, though with important inconsistencies across datasets and tasks \cite{hua2025large}. For example, elevated negative valence and absolutist wording have been observed in language linked to depression and anxiety, but they are not specific to any single disorder \cite{almosaiwi2018absolutist, nlp24_survey}. Temporal focus provides clinical context: past-oriented narratives can align with trauma “re-experiencing” content (an intrusion cluster in DSM-5-TR), whereas present-focused negative self-evaluations are consistent with rumination in depressive episodes; again, these are cues rather than determinations \cite{edition2013diagnostic}. In our algorithm, these features are categorical inputs in $\mathcal{E}$ that help organize evidence and can modulate override priors ($b_{\uparrow}, b_{\downarrow}$), while the interface renders them as continuous ribbons so clinicians can skim (System~1) and drill down to quotes (System~2) without over-interpreting any single marker.

\subsection*{Step 3: Structured excerpts as theory-aligned evidence}
\begin{figure}[t]
  \centering
  \includegraphics[width=1.00\linewidth]{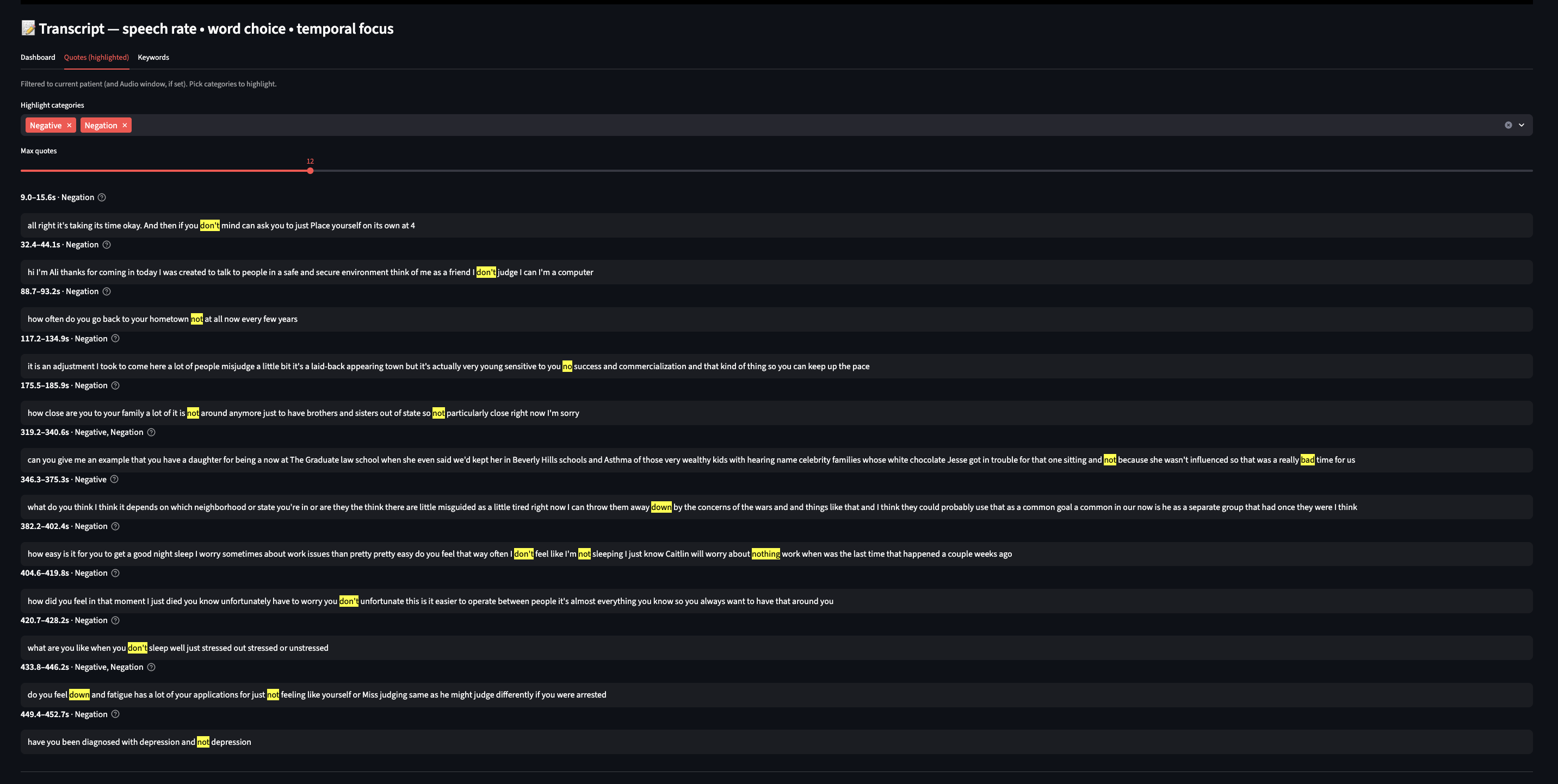}
  \caption{\textbf{Quotes view.} Clinicians can filter utterances by diagnostic categories, surfacing concise evidence for or against AI suggestions.}
  \label{fig:transcript-quotes}
\end{figure}

While ribbons provide a macro-level index, clinicians require micro-level evidence to evaluate or contest AI suggestions. The quotes panel (Fig.~\ref{fig:transcript-quotes}) surfaces utterances that match theory-aligned categories of avoidance, hyperarousal, and negative self-statements highlighted and timestamped. This is a design response to the risk of automation bias: instead of passively accepting the model’s severity estimate, clinicians are invited to inspect a curated but transparent subset of the transcript (the patient's answers only). Algorithmically, these quotes are embedded in the log as text spans linked to timestamps, ensuring reproducibility. This approach resonates with findings in explainable AI that curated, task-relevant exemplars foster both trust and accountability \cite{sivaraman2023ignore,lee2021human}. By enabling rapid access to counterevidence, the system operationalizes a disconfirmatory search strategy, which is critical in high-stakes diagnostic reasoning.

\subsection*{Step 4: Contrastive keyword analysis}
\begin{figure}[t]
  \centering
  \includegraphics[width=\linewidth,trim=0 0 0 6,clip]{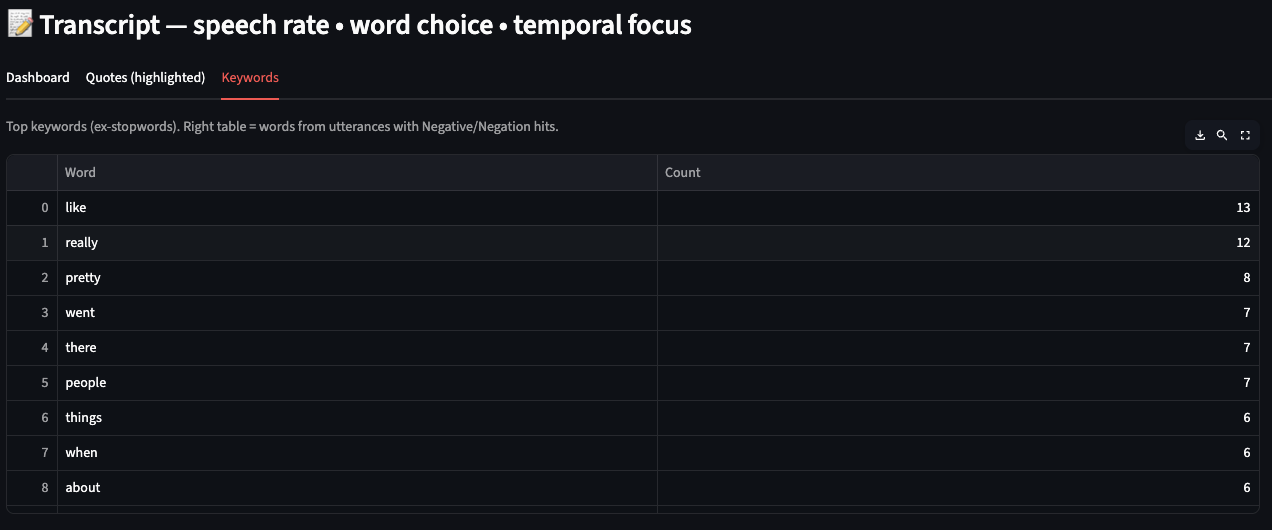}\vspace*{-1mm}
  \includegraphics[width=\linewidth,trim=0 6 0 0,clip]{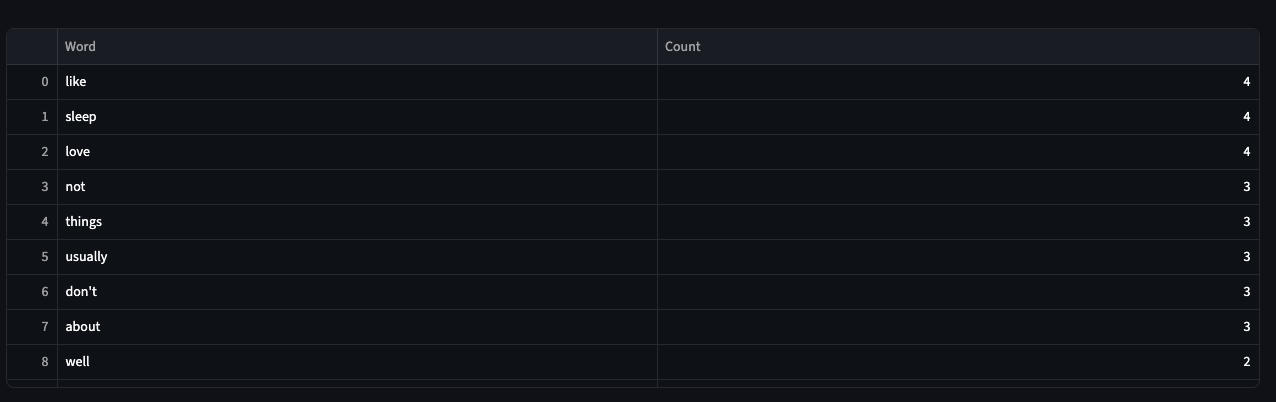}
  \caption{\textbf{Keyword contrast tables.} Left: top keywords for the full session.
  Right: keywords restricted to negative/negated utterances, highlighting contrastive salience.}
  \label{fig:transcript-keywords}
\end{figure}

The final transcript-related panel (Fig.~\ref{fig:transcript-keywords}) offers a contrastive lens: two tables, one showing globally salient keywords, and the other restricted to negative or negated contexts. This comparative framing draws on structural-alignment theory in cognitive science
\cite{gentner2001structural}:
clinicians tend to privilege explanations that account for discrepant
distributions when alternatives are compared and aligned. In practice, if terms such as ``sleep’’ or ``don’t’’ disproportionately cluster in negative contexts, they may corroborate DSM clusters of insomnia or anhedonia. Conversely, their absence may support a downward override. Again, these distributions are included in $\mathcal{E}$ and carried into logs, ensuring that downstream analyses of override rates can be traced back to contrastive evidence. The design aligns with HCI principles of parsimony and transparency: instead of opaque embeddings, the system presents interpretable contrasts that allow clinicians to reason about alignment between language use and diagnostic categories.

\subsection*{Step 5: Embodied face and gaze signals}
\begin{figure}[t]
  \centering
  \includegraphics[width=1.00\linewidth]{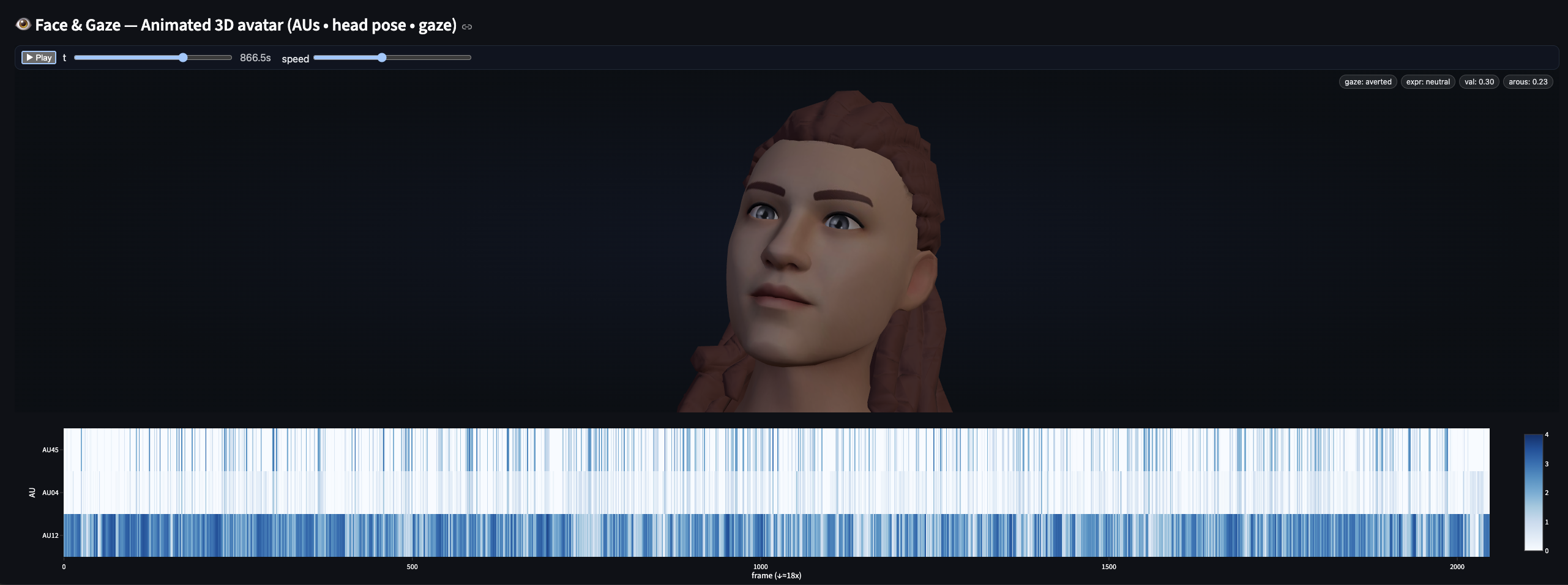}
  \vspace{6pt}
  \includegraphics[width=1.00\linewidth]{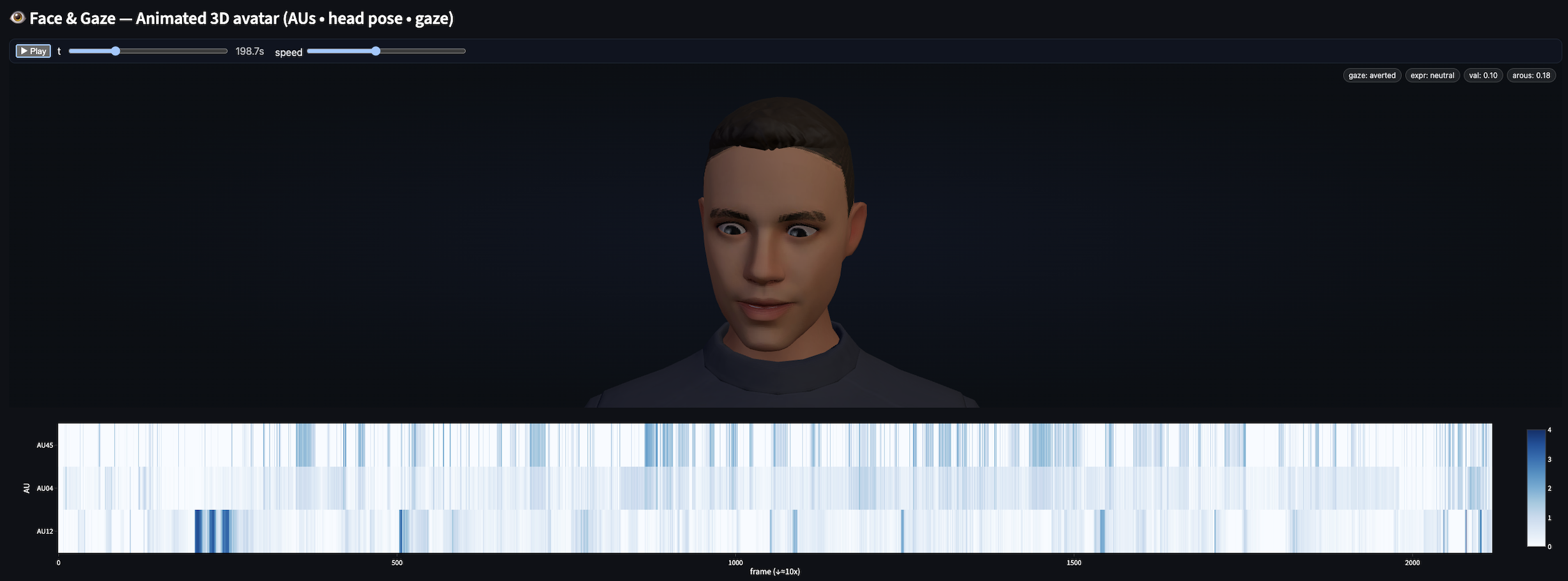}
  \caption{\textbf{Face \& Gaze } Replayable avatars synchronized to OpenFace features, with AU heatmaps and gaze/expression state chips. }
  \label{fig:facegaze-top}
\end{figure}

The next stage of the workflow brings facial action units and gaze dynamics into the clinician’s perceptual field. The innovation here is not only the inclusion of AU heatmaps, but the use of an avatarization pipeline: OpenFace features (AU12, AU04, AU45, gaze vectors) drive a neutral animated avatar that mirrors facial expressions and gaze orientation. Unlike raw video playback, this abstraction preserves the ecological validity of nonverbal cues while removing biometric identifiers, enabling researchers to scrutinize patterns of affective blunting, gaze aversion, or engagement without exposing the patient’s identity. In practice, the avatarization augments the AU overlay (Step~6) by providing both a perceptual and quantitative lens on the same signals.
 As shown in Fig.~\ref{fig:facegaze-top}, SimClinician presents an animated avatar synchronized with the OpenFace streams, where AU12 (smile), AU04 (brow tension), and AU45 (blink) are continuously plotted as heatmaps.  Compact state chips report gaze direction, coarse valence, and arousal shifts, maintaining consistency with the encodings introduced in the transcript and audio panels. The design choice to employ a replayable avatar rather than static plots is intentional: it reflects ecological interface design principles, making engagement and affective flatness perceptible at a glance while retaining access to raw signal traces. This dual layer immediacy plus traceability prevents premature closure, a known cognitive risk in high-stakes clinical reasoning \cite{asan2020artificial}. Algorithmically, the AU and gaze traces are added to the evidence set $\mathcal{E}$; their influence on overrides is modulated by priors $b_{\uparrow}, b_{\downarrow}$ when clinicians weigh blunted affect or gaze aversion against self-report anchors. In this way, the face–gaze module becomes not just a visualization but a driver of policy-sensitive dynamics in Algorithm~\ref{alg:simclinician}.

\subsection*{Step 6: From framewise signals to rule-based streaks}
\begin{figure}[t]
  \centering
  \includegraphics[width=1.00\linewidth]{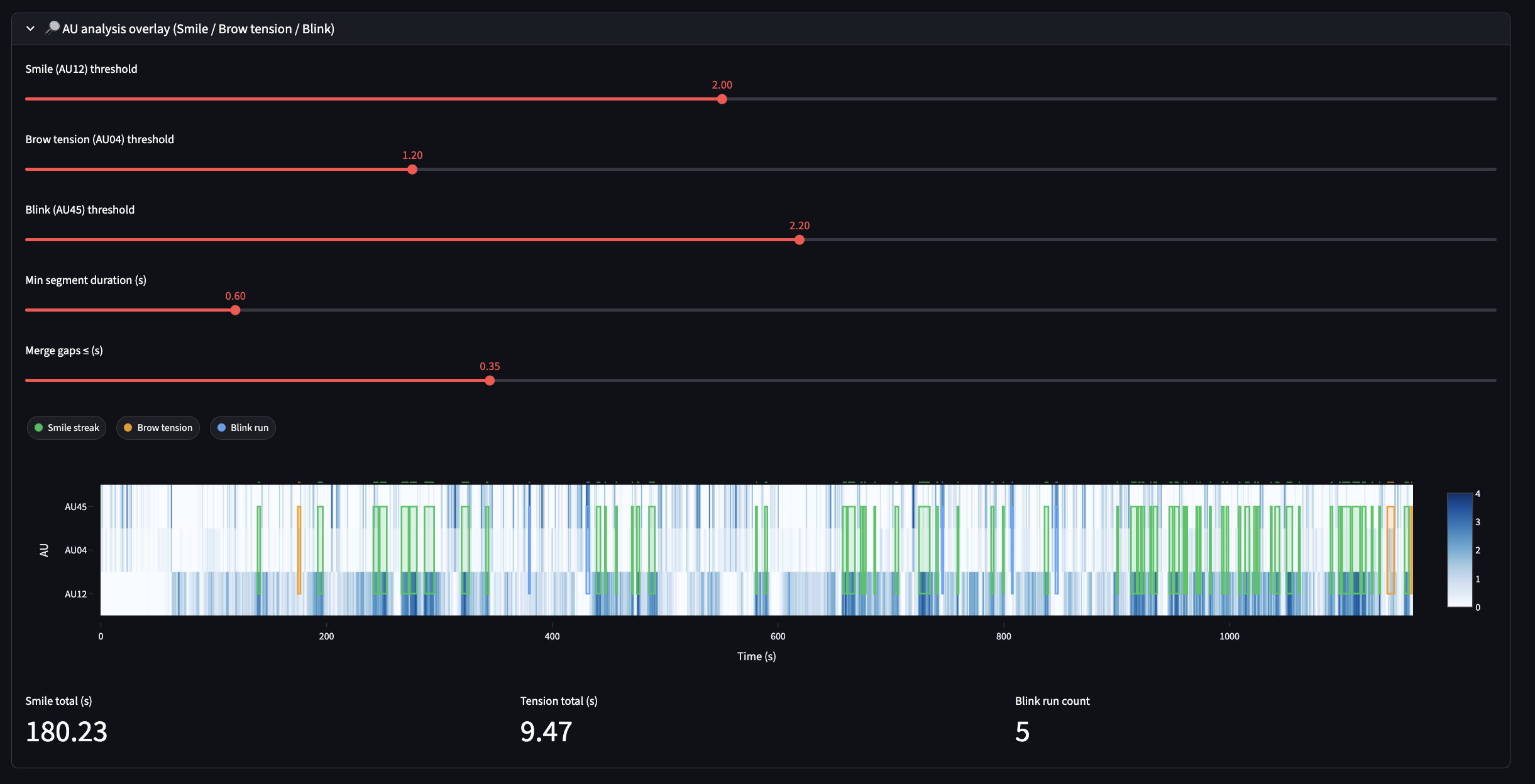}
  \caption{\textbf{AU overlay.} Rule-based grouping converts noisy AU traces into contiguous intervals: smiles (green), tension runs (orange), and blink bursts (blue). Totals are tallied and carried forward into the fusion step.}
  \label{fig:facegaze-overlay}
\end{figure}

Raw AU streams can be noisy and difficult to interpret across thousands of frames. To address this, SimClinician overlays interactive, rule-based detectors that group contiguous intervals into interpretable “runs” of smiles, tension, and blinks (Fig.~\ref{fig:facegaze-overlay}). Threshold sliders, minimum-duration controls, and merge-gap parameters externalize the assumptions clinicians routinely make informally about what counts as a genuine expression versus noise. By exposing these parameters, the design operationalizes the HCI principle of parameter probing, inviting clinicians to test whether patterns survive stricter criteria. This transparency is not cosmetic: the streak totals are exported as structured features into the fusion step, ensuring parity between subjective probing and logged evidence. Within Algorithm~\ref{alg:simclinician}, these features contribute to the risk score $R$ and can tilt override likelihoods depending on policy priors. In effect, the overlay makes the “black box” of computer vision legible in a form that clinicians can contest, calibrate, and trust.

\subsection*{Step 7: Fusion and decision as judge–advisor negotiation}
\begin{figure}[t]
  \centering
  \includegraphics[width=1.00\linewidth]{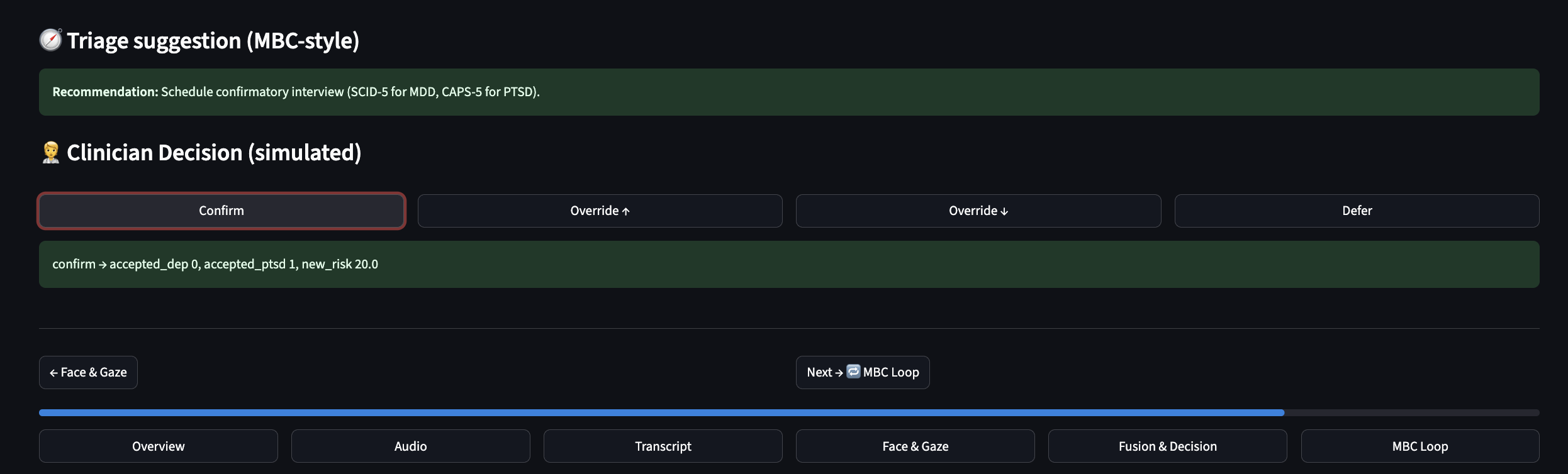}
  \caption{\textbf{Fusion \& Decision.} A banded risk gauge aggregates multimodal evidence into a severity score. The action bar offers \emph{Confirm}, \emph{Override} (up/down), or \emph{Deferral}, each logged with full contextual metadata.}
  \label{fig:fusion-decision}
\end{figure}

The culmination of the workflow is the decision screen (Fig.~\ref{fig:fusion-decision}), where multimodal evidence, AI prediction, and clinician judgment converge. Here, the algorithmic recommendation is presented not as a binary label but as a risk gauge with color-banded zones and plain-language next-step suggestions (e.g., “Schedule confirmatory interview”). This framing explicitly encodes uncertainty and downstream actionability, aligning with HCI research that emphasizes utility-focused rather than purely probabilistic outputs \cite{bhatt2023collaborative,asan2020artificial}. Crucially, this is the point where Algorithm~\ref{alg:simclinician} enforces soft friction: confirmation requires only a single click, but overrides trigger a lightweight attestation step, operationalizing default-effect theory to dampen impulsive changes without removing autonomy. All contextual evidence, thresholds, and the chosen action $a$ are written to the log $\mathcal{L}$, ensuring that every decision is auditable and replayable. In this way, the decision panel instantiates a judge–advisor setting: the AI serves as an advisor offering an initial recommendation, and the clinician—as judge—assigns weight to that advice, with light confirmation friction and explanations designed to foster \emph{appropriate reliance} rather than over- or under-use of the AI \cite{Bailey2023WeightOfAdvice,Schemmer2023AppropriateReliance}.

\subsection*{Step 8: Closing the loop with longitudinal tracking}
\begin{figure}[t]
  \centering
  \includegraphics[width=1.00\linewidth]{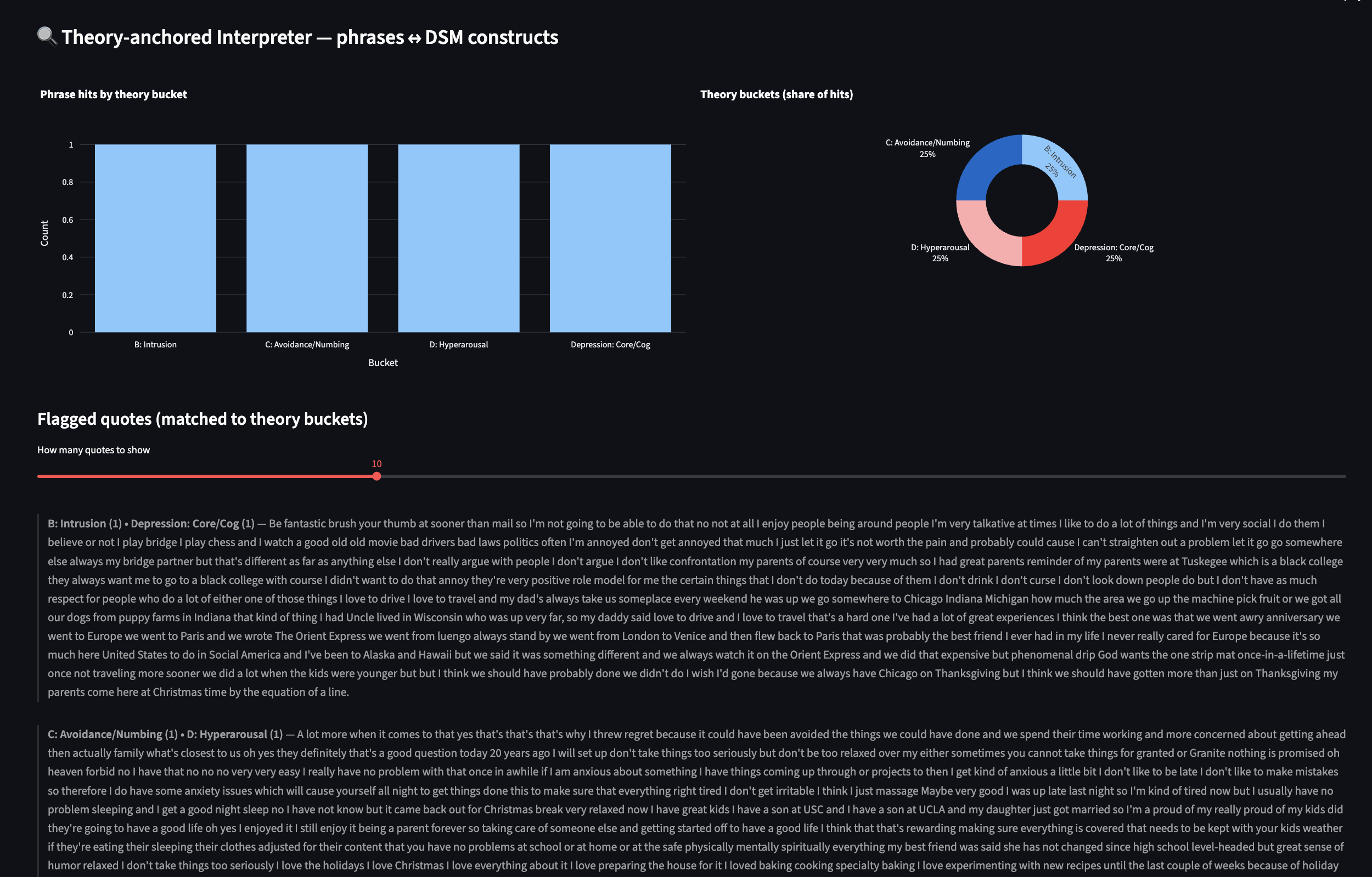}
  \caption{\textbf{Measurement-Based Care loop.} Longitudinal comparisons flag reliable and clinically significant change, linking diagnostic decisions to ongoing trajectories of care.}
  \label{fig:mbc}
\end{figure}

Unlike static benchmark models, clinical care unfolds over-repeated encounters. SimClinician therefore ends the workflow with an MBC loop (Fig.~\ref{fig:mbc}), where longitudinal PHQ and PCL scores are tracked and compared against validated thresholds of reliable and clinically significant change \cite{blampied2022reliable,marx2022reliable}. Risk-band crossings and deltas are explicitly annotated, enabling clinicians to recalibrate treatment intensity based on evidence of progress or deterioration. Algorithmically, these deltas update metadata $M$ for subsequent sessions, influencing baseline thresholds $\tau_d, \tau_p$ in the policy $\pi$. The design closes the loop from scan to track, aligning SimClinician not only with HCI principles of auditability and transparency but also with clinical standards of iterative, evidence-based care \cite{trivedi2020implementing,dams2023measurement}. By embedding longitudinal logic directly into both UI and controller, the system demonstrates that diagnostic reliability is not a one-off event but a trajectory.

\subsection*{Summary of workflow design}
Taken together, the eight steps embody a coherent design philosophy. \emph{The order shown is expository, not prescriptive:} the dashboard supports non-linear use, so a clinician may begin with the transcript, audio, or face/gaze panels and shuttle between them; all views are synchronized by a shared timeline, and the controller/logs are order-agnostic and record the actual traversal. Multimodal signals are surfaced sequentially, each grounded in established diagnostic constructs; parameters are exposed for probing to calibrate trust; algorithmic policies translate cognitive biases into measurable distributions; and all actions are logged to support reproducibility and audit. The algorithm provides the skeleton, while the interface panels flesh out perceptual and reflective affordance. This choreography \emph{is designed to} reduce cognitive load, encourages disconfirmatory reasoning, and ensures that clinician authority is preserved even as AI recommendations shape the decision space. By bridging psychiatric gold standards with HCI principles, SimClinician establishes itself as more than a dashboard: it is a methodological harness for studying clinician–AI collaboration with rigor, scalability, and accountability.\textit{We state these as design goals rather than outcomes; quantitative proxies for effort and responsiveness (accept/override/deferral mix and decision latency) are reported in Section~\ref{sec:evaluation}.}

\section{System Evaluation}\label{sec:evaluation}

\noindent The goal of our evaluation is not to show incremental gains in predictive accuracy but to demonstrate how \systemname{} enables systematic, reproducible exploration of the interactional space between clinicians and AI. In line with the methodological framing of Section~\ref{sec:workflow}, we structure the evaluation around three levers that have been foregrounded in decision science and HCI research as decisive for adoption and reliability: the insertion of micro–cost frictions on overrides, the shaping of collaboration styles through policy priors, and the modulation of cognitive load through time budgets \cite{asan2020artificial,bhatt2023collaborative,smith2024clinicians}. Our experiments exercise the simulation harness across all combinations of these levers and interface design factors, enabling us to move from anecdotal observations to principled characterizations of behavior at scale.

\paragraph{Experimental setup.}
We ran a full factorial experiment across five dimensions of the interface and policy space: \emph{display format} (numeric vs.\ banded), \emph{explanations} (off vs.\ on), \emph{override friction} (none vs.\ confirmation), \emph{time budget} (short vs.\ long), and \emph{policy priors} (safety, parsimony, deferral). Each cell was populated with $10{,}000$ simulated cases drawn from E-DAIC, producing 48 cells in total and hundreds of thousands of logged interactions. Because the controller enforces parity between UI and API modes, the same codepath generated both the interactive trials and the large batch runs, guaranteeing reproducibility. All figures in this section are generated directly from these logs. We report outcomes in absolute terms with Wilson 95\% confidence intervals, ensuring both clarity and statistical robustness.

\paragraph{Psychological framing of the levers.}
Each lever corresponds to a well-documented cognitive or social mechanism. \emph{Override friction} operationalizes default and omission effects: a small required click can stabilize choices by making deviation from a recommendation non-trivial, reflecting bounded rationality in time-constrained environments \cite{liang2022save}. \emph{Policy priors}  consistent with this framing, we design the UI and policies to promote appropriate reliance and end-to-end team performance: clinicians can confirm confident recommendations and override or deferral when evidence is weak—an approach aligned with human-AI teaming optimized for team utility rather than model accuracy \cite{bansal2021most} and with clinical guidance to calibrate (not maximize) trust in AI \cite{asan2020artificial}. Finally, \emph{time pressure} modulates the reliance on fast heuristics over slow deliberation, amplifying confirmation bias and satisfying tendencies \cite{asan2020artificial, bansal2021most}. By embedding these constructs directly in Algorithm~\ref{alg:simclinician}, we treat cognitive dynamics as first-class objects of study rather than uncontrolled noise.

Our evaluation proceeds in three stages that mirror the cognitive and design levers foregrounded in Section~\ref{sec:workflow}. First, we ask whether lightweight frictions modeled as confirmation clicks---meaningfully stabilize decision distributions without impeding flow. Second, we examine how the three policy priors defined in Section~\ref{sec:workflow} and instantiated in Algorithm~\ref{alg:simclinician} 
\texttt{safety} (upward-leaning; escalate when in doubt),
\texttt{parsimony} (downward-leaning; avoid decsion escalation when strong evidence is absent ),
and \texttt{deferral} (prefer additional information/second opinion)—
shape collaboration styles from alignment to skepticism to deliberation.

\begin{figure}[t]
  \centering
  \includegraphics[width=\figwidth]{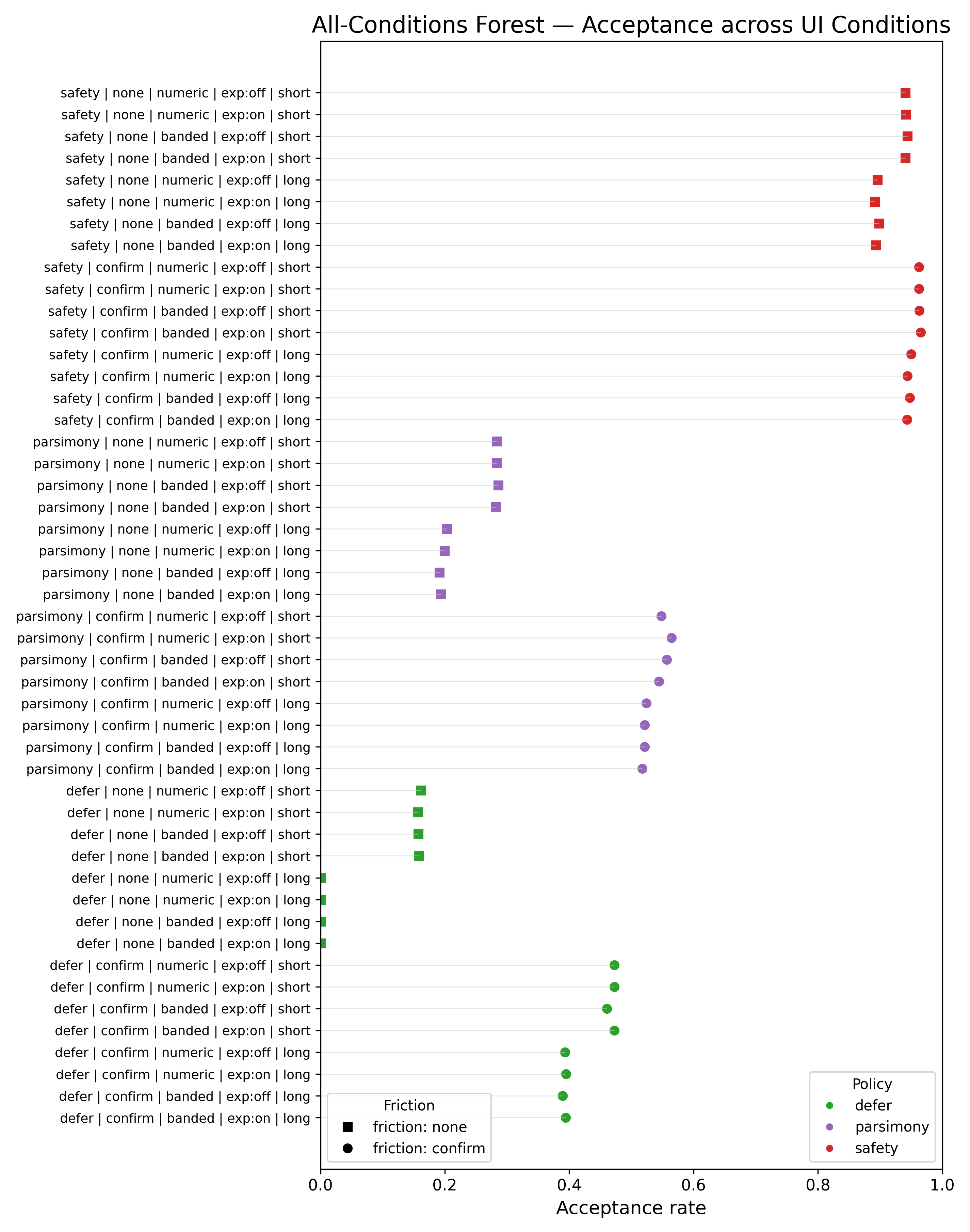}
  \caption{\textbf{Acceptance rates across \emph{every} UI condition.}
  Rows enumerate \texttt{policy} \textbar\ \texttt{friction} \textbar\ \texttt{display} \textbar\ \texttt{explanations} \textbar\ \texttt{time}.
  Color encodes policy (green=\texttt{deferral}, purple=\texttt{parsimony}, red=\texttt{safety}); marker shape encodes friction (circle=\texttt{confirm}, square=\texttt{none}).}
  \label{fig:forest}
\end{figure}

\paragraph{Global acceptance structure.}
We begin with the broad question: does override friction matter? The forest plot in Fig.~\ref{fig:forest} provides a global answer by summarizing acceptance across all UI$\times$policy cells. The \texttt{deferral} prior distributes choices across \emph{accept}, \emph{override}, and \emph{deferral}, producing the most heterogeneous outcome structure. Most importantly, introducing a single confirmation click shifts points rightward across policies, raising acceptance and lowering overrides. This pattern is consistent with default/omission effects—when the cost of deviating from a recommendation is raised even minimally, alignment with the AI becomes the path of least resistance \cite{liang2022save,ezran20231190}. Rather than an incidental UI quirk, this shows how the system translates cognitive theory into measurable system-level distributions, linking micro-level frictions to macro-level patterns of clinician–AI negotiation that can be reproduced and analyzed. 

At a glance, Fig.~\ref{fig:forest} shows three regularities: \texttt{safety} yields the highest acceptance across UI/time cells; \texttt{parsimony} yields the lowest acceptance (more overrides$\downarrow$); and enabling \texttt{confirmation} increases acceptance within every policy—the corresponding points lie to the \emph{right} of their no-friction counterparts.

\begin{figure}[t]
  \centering
  \includegraphics[width=.70\linewidth]{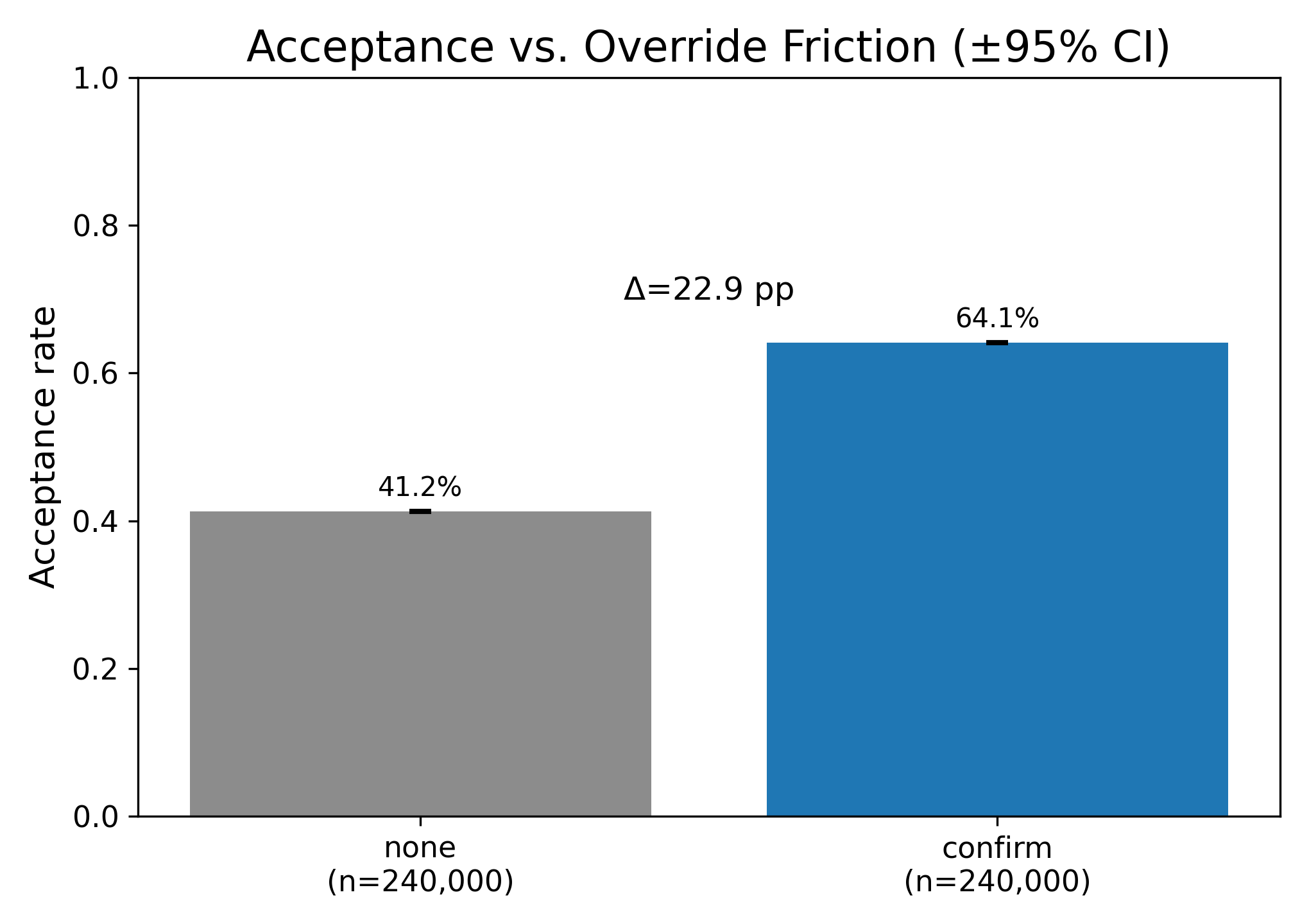}
  \caption{\textbf{Friction sharply increases acceptance.}
  Bars show the overall acceptance rate under \texttt{none} vs.\ \texttt{confirm} with Wilson 95\% CIs (n shown under each bar).
  The absolute difference is roughly $+22.9$ percentage points.}
  \label{fig:acceptance-friction}
\end{figure}

\paragraph{Magnitude of the friction effect.}
The global pattern is quantified in Fig.~\ref{fig:acceptance-friction}, which compares acceptance rate under no friction versus confirmation friction. Requiring a single additional click raises acceptance by +22.9 percentage points, a strikingly large effect given the minimal change in interface mechanics. This demonstrates that the system captures not only micro-level UI tweaks but also their macro-level impact on clinical reliability. In a live deployment, such a change could spell the difference between widespread override of AI suggestions and a culture of calibrated acceptance. The scale of the effect aligns with prior work on attestation prompts in EHRs, where small interface frictions reduced inappropriate prescribing by double-digit percentages \cite{ezran20231190}. \systemname{} generalizes this phenomenon into a reusable testbed for tuning override frictions before exposing them to clinicians.

\begin{figure}[t]
  \centering
  \includegraphics[width=\figwidth]{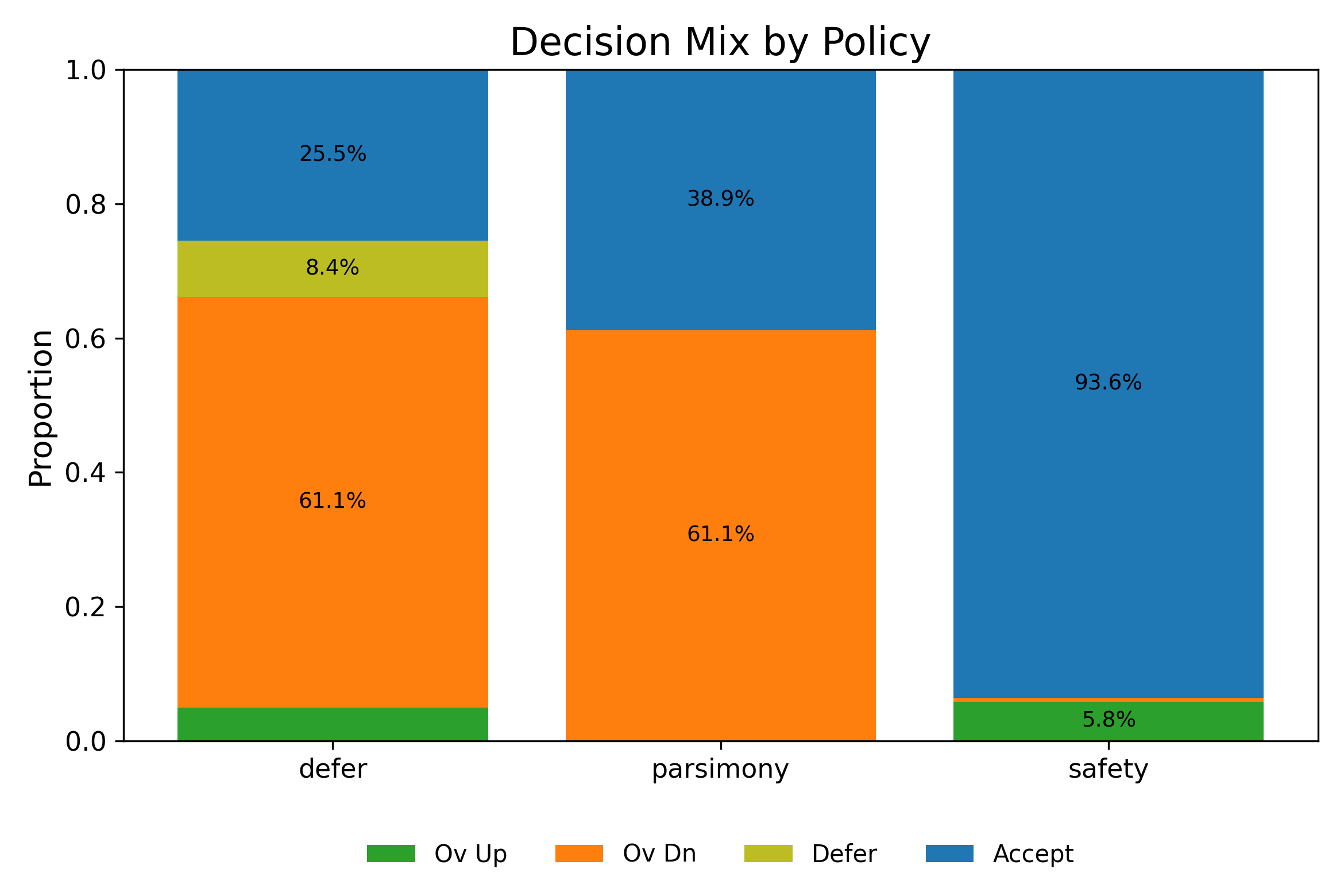}
  \caption{\textbf{Decision Mix by Policy.}
  Stacked proportions of outcomes aggregated across UI/time cells.
  \texttt{Safety} is dominated by \emph{Accept} ($\approx$94\%) with a small \emph{Override↑};
  \texttt{parsimony} splits between \emph{Override↓} ($\approx$61\%) and \emph{Accept} ($\approx$39\%);
  \texttt{deferral} shows a mixed profile with substantial \emph{Override↓} and some \emph{Deferral}.}
  \label{fig:policy-mix}
\end{figure}

\paragraph{Collaboration styles under different priors.}
Having established that friction systematically raises acceptance, we next ask how collaboration priors steer decision mix. Fig.~\ref{fig:policy-mix} disaggregates the outcomes by policy, showing how each institutional stance translates into observable override patterns.
 Under \texttt{safety}, acceptance dominates at $\approx94\%$, with a small fraction of upward overrides. Under \texttt{parsimony}, by contrast, downward overrides dominate at $\approx61\%$, reflecting the prior’s built-in aversion to escalation. The \texttt{deferral} of prior shows a heterogeneous distribution with substantial down-overrides and a visible deferral fraction, offering an institutional option where deliberation and second opinions are encouraged. By enabling these priors to be parameterized and tested, the system creates a bridge between abstract organizational values (e.g., “erring on the side of safety”) and their measurable behavioral consequences at the decision level.

\begin{figure}[t]
  \centering
  \includegraphics[width=.65\linewidth]{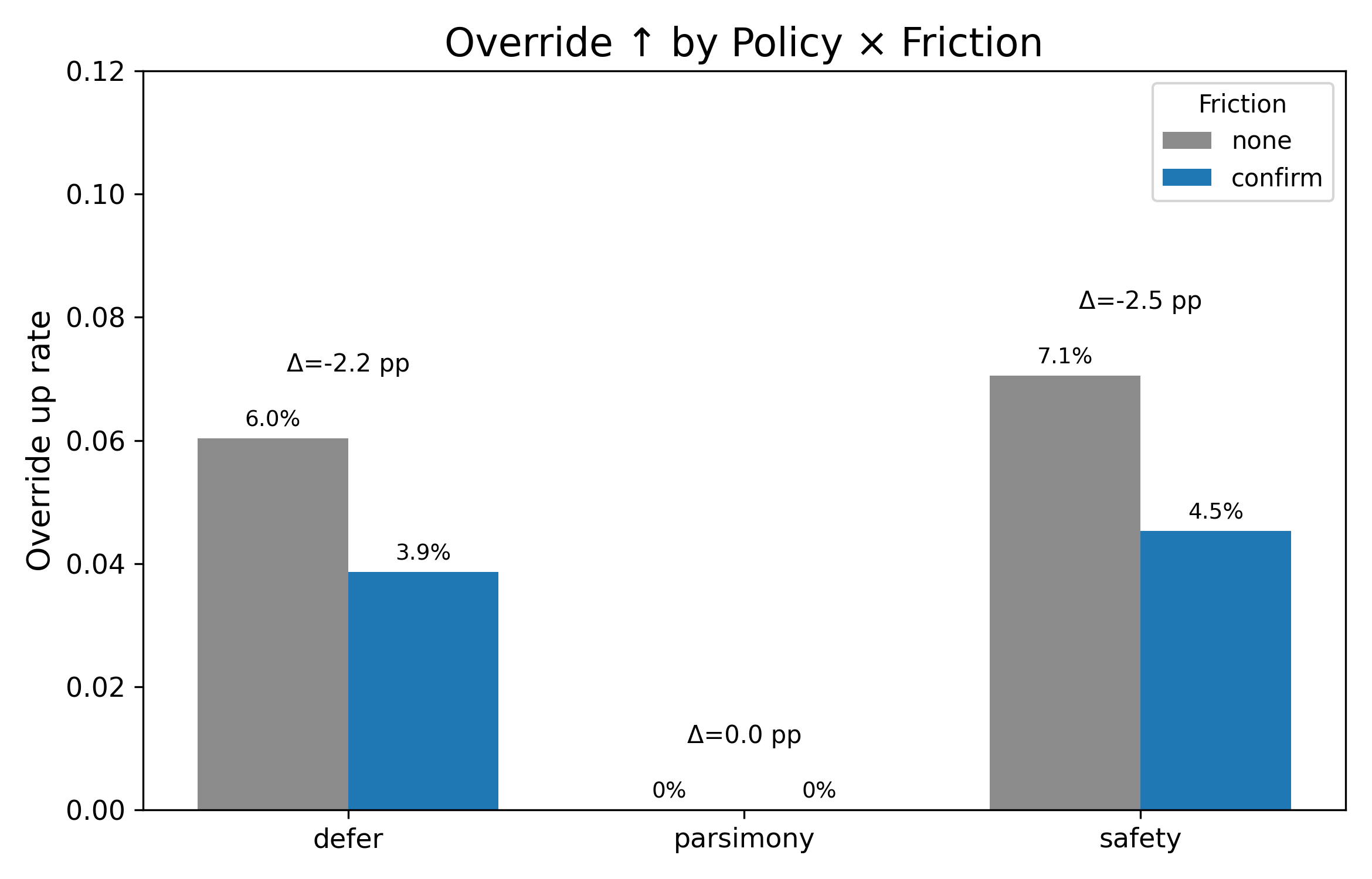}
  \caption{\textbf{Override $\uparrow$ by policy$\times$friction (grouped bars).}
  Each policy shows two bars: \texttt{none} (grey) and \texttt{confirm} (blue).
  Soft friction reduces upward overrides by $\approx$2--3\,pp in \texttt{safety} and \texttt{deferral}; \texttt{parsimony} stays near zero (by construction).}
  \label{fig:override-grouped}
\end{figure}

\paragraph{Targeting escalation risks.}
A particularly important finding concerns upward overrides, which escalate care even when the AI does not recommend it. Fig.~\ref{fig:override-grouped} shows how soft confirmation friction trims these cases. In both the \texttt{safety} and \texttt{deferral} priors, upward overrides drop by 2–3 percentage points when confirmation is required, while \texttt{parsimony} stays near zero by design. Even in the worst-case cell \texttt{safety} with numeric display, explanations on, no friction, and long time upward overrides remain below 9\%. These results are clinically meaningful: unnecessary escalation is expensive in terms of specialist referrals, patient anxiety, and resource allocation. By bounding escalation risk through lightweight interface design, \systemname{} provides an actionable lever for balancing safety and efficiency.

\begin{figure}[t]
  \centering
  \includegraphics[width=.65\linewidth]{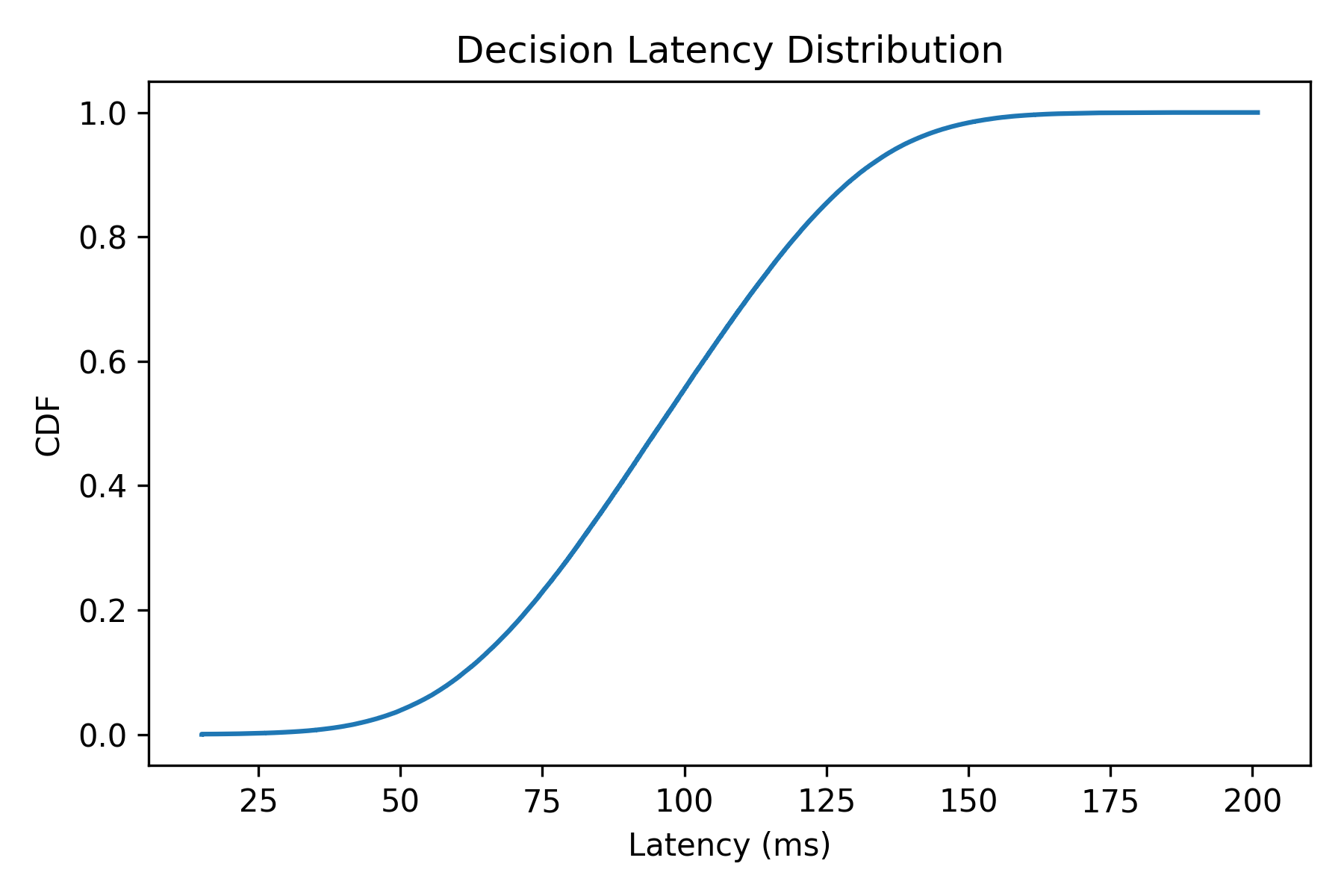}
  \caption{\textbf{Decision latency distribution.}
  The empirical CDF shows decisions concentrate well under 150\,ms; 95th percentile $\approx$139\,ms.}
  \label{fig:latency}
\end{figure}

\paragraph{Responsiveness and cognitive cost.}
Finally, any design intervention must be judged not only by its effect on reliability but also by its cost to responsiveness. If friction buys safety at the price of sluggish interaction, it fails as a practical design choice. Any intervention that raises acceptance rates must be carefully weighed against its impact on responsiveness. Clinical workflows are highly time-sensitive, and additional micro–costs can quickly accumulate into perceptible slowdowns. Fig.~\ref{fig:latency} plots the empirical cumulative distribution of decision latencies across all conditions. The results are striking: even with confirmation friction enabled, decisions cluster well below 150\,ms, with the 95th percentile at approximately 139\,ms. This value sits comfortably below thresholds where human–computer interaction studies have shown latency to become disruptive in high-stakes domains \cite{asan2020artificial,bhatt2023collaborative}. In other words, the friction effect that so strongly shifts acceptance distributions does not introduce a meaningful performance penalty. When contextualized against the 23 percentage point acceptance gain, this trade-off clearly favors adoption.

\begin{figure}[t]
  \centering
  \includegraphics[width=\figwidth]{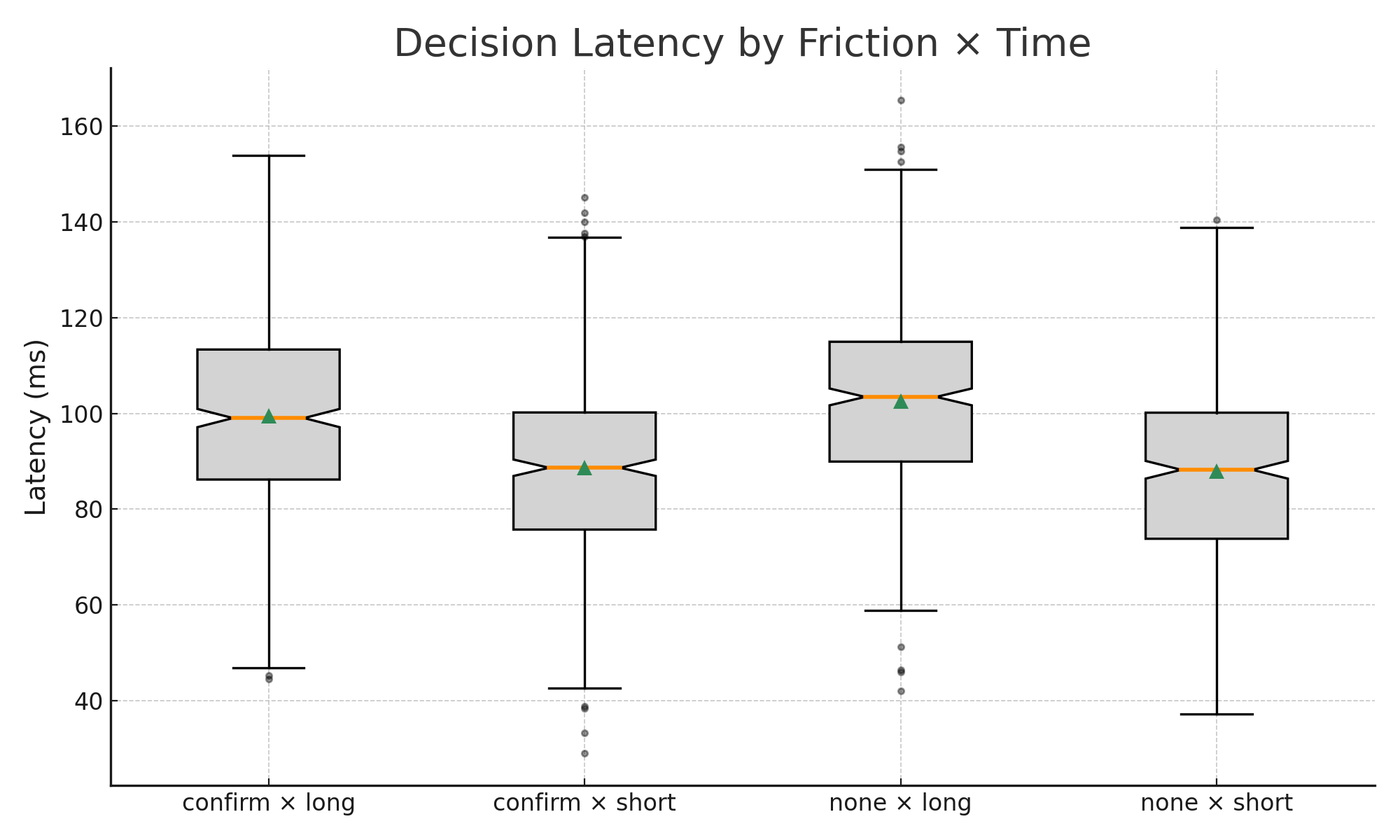}
  \caption{\textbf{Decision latency by Friction $\times$ Time.}
  Notched boxplots show the median (notch), IQR (box), whiskers, outliers, and the mean (triangle).
  \texttt{confirm} introduces a small latency premium relative to \texttt{none}, while \texttt{short} budgets reduce latency across both friction settings.}
  \label{fig:latency-fx-time}
\end{figure}

To further disentangle whether the latency distribution is shaped by time pressure or by friction itself, Fig.~\ref{fig:latency-fx-time} crosses the two factors. The boxplots show that confirmation friction does add a modest latency premium relative to no friction, but the shift is small and consistent. In contrast, short time budgets compress latency across both friction settings, highlighting how time constraints override micro–costs in shaping decision pace. Taken together, the figures demonstrate that soft confirmation frictions do not meaningfully degrade responsiveness, even under pressure, and thus represent a low-cost, high-yield intervention in clinician–AI interaction.

In summary, the first three analyses build a coherent picture: (i) soft confirmation frictions dramatically stabilize acceptance; (ii) collaboration priors encode institutional philosophies with measurable consequences; and (iii) these levers do not degrade responsiveness even under time pressure. We now turn to a complementary concern: whether the underlying probabilities themselves are trustworthy and communicable.

\paragraph{Probability calibration and risk communication.}
Beyond interactional levers, deployment of predictive systems requires calibrated probabilities that clinicians can interpret and act upon. Reliability diagrams and scalar metrics such as Expected Calibration Error (ECE) and Maximum Calibration Error (MCE) are canonical for this purpose \cite{arrietaibarra2022metricscalibrationprobabilisticpredictions}, though recent work highlights their sensitivity to binning and sample size, advocating smoother kernel-based approaches with stronger theoretical guarantees \cite{blasiok2023unifying}. Clinical adoption, however, demands an even broader perspective: calibration must be evaluated not only statistically but in terms of decision utility. Decision Curve Analysis (DCA) exemplifies this shift, weighting true versus false positives to compute net benefit across thresholds \cite{sadatsafavi2021moving,netto2024bayesian,zhao2024understanding,mijderwijk2021my}. Risk communication is equally critical: static calibration plots often obscure subgroup heterogeneity, whereas interactive visualization tools such as \emph{Calibrate} allow clinicians to drill into subgroups, improving interpretability and trust \cite{xenopoulos2022calibrate}.

\begin{figure}[t]
  \centering
  \includegraphics[width=0.65\linewidth]{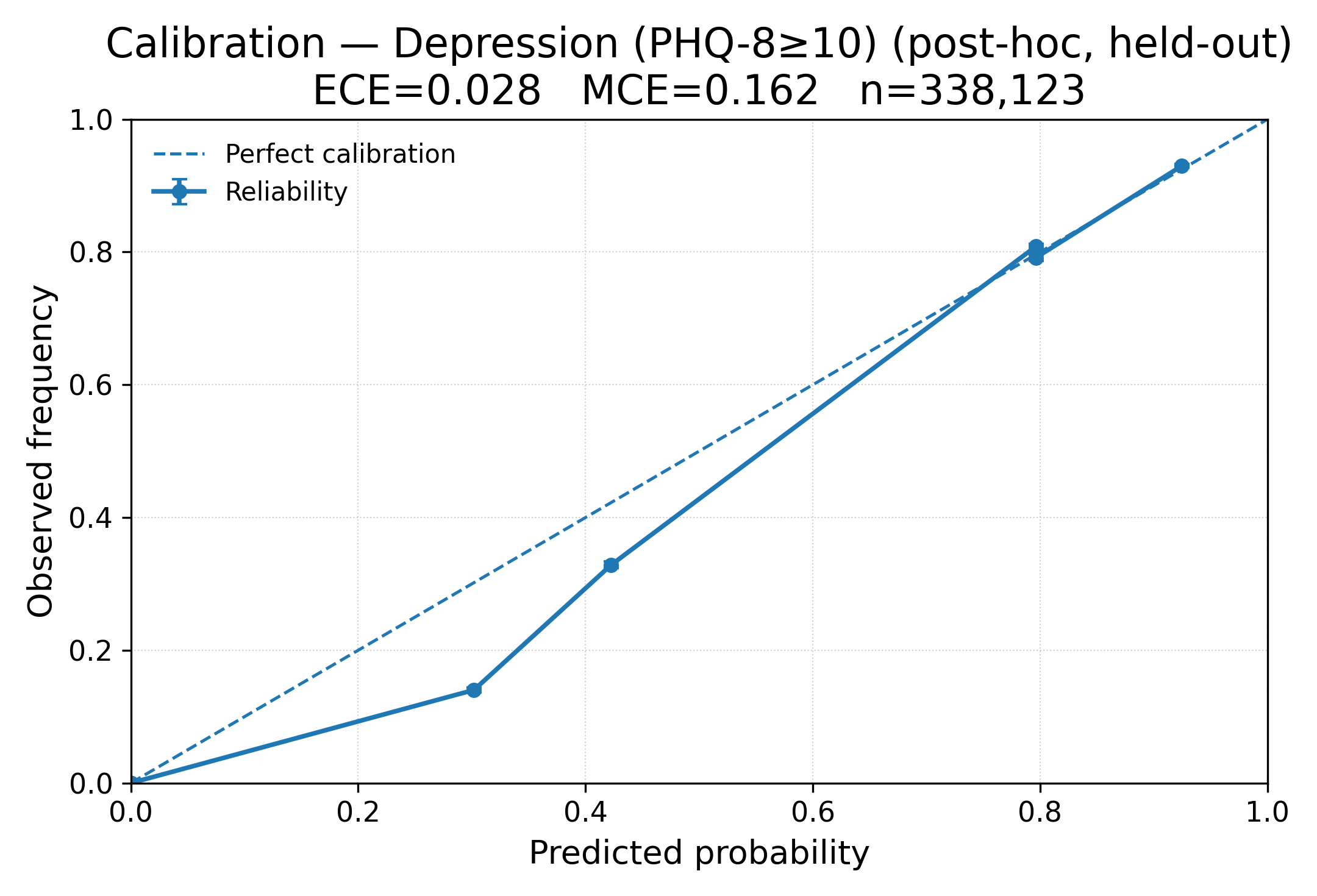}
  \caption{Calibration curve for depression (PHQ-8$\geq$10) after post-hoc isotonic regression on held-out validation data. Reliability is strongly improved (ECE=0.028, MCE=0.162, $n=338{,}123$).}
  \label{fig:calib_dep}
\end{figure}

\begin{figure}[t]
  \centering
  \includegraphics[width=0.67\linewidth]{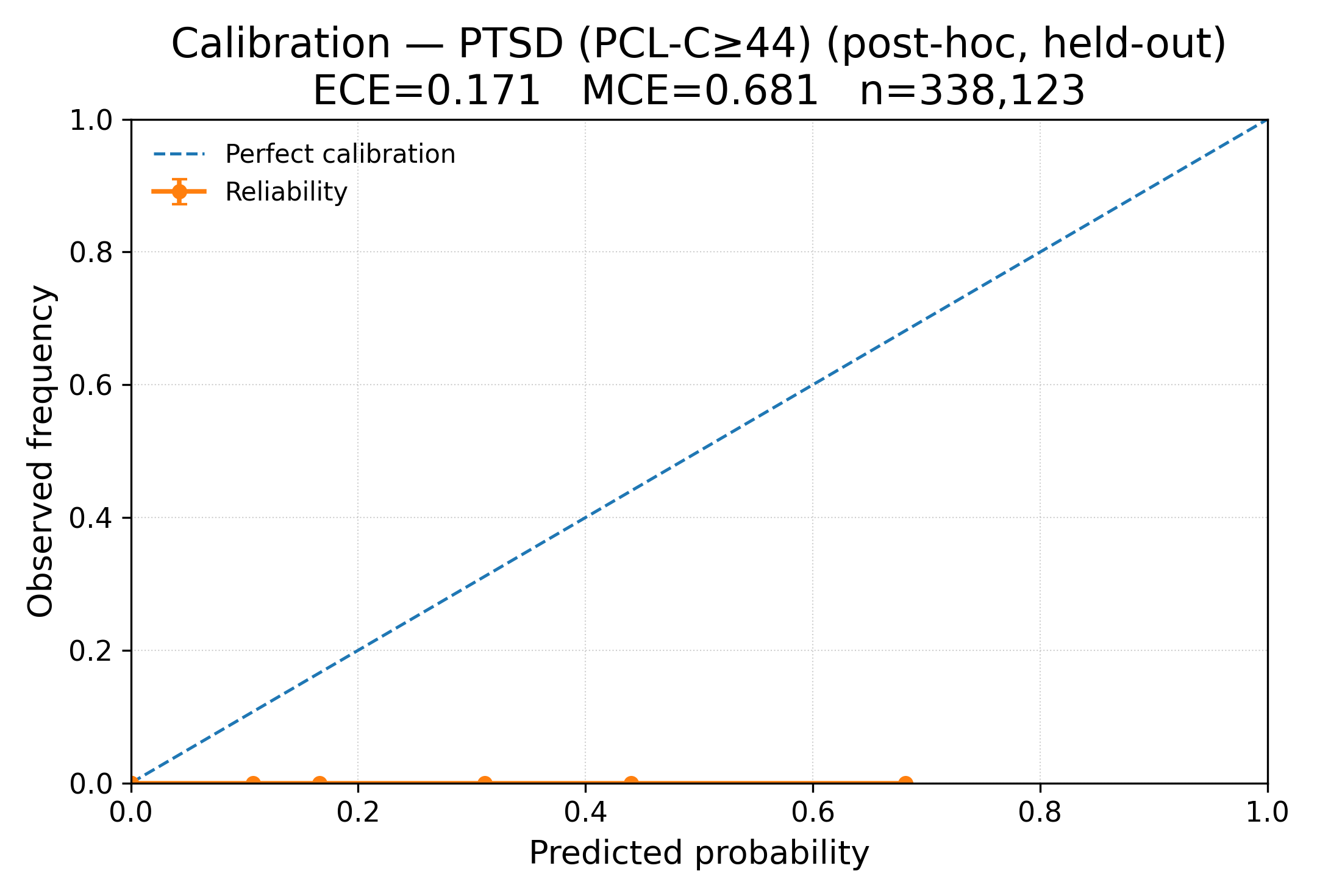}
  \caption{Calibration curve for PTSD (PCL-C$\geq$44). Despite post-hoc isotonic regression, predictions remain poorly calibrated (ECE=0.171, MCE=0.681, $n=338{,}123$) due to severe class imbalance.}
  \label{fig:calib_ptsd}
\end{figure}

Within \systemname{}, we applied post-hoc calibration to model outputs using a held-out participant split. Thirty percent of participants were used exclusively to fit a monotonic mapping (Isotonic Regression with out-of-bounds clipping), while the remaining 70\% were reserved for evaluation. In conditions where PTSD positives were insufficient, we fell back to Platt scaling. Importantly, calibration does not alter discrimination (AUC/F1) but adjusts the reliability of probabilities. Figs.~\ref{fig:calib_dep} and~\ref{fig:calib_ptsd} report calibration curves for depression and PTSD respectively. For depression, isotonic regression substantially reduced miscalibration: ECE dropped from 0.128 to 0.028 and MCE from 0.501 to 0.162, producing a curve that closely follows the diagonal. This improvement ensures that clinicians can trust mid-range risk estimates, critical for triage decisions. In contrast, PTSD calibration remains poor (ECE=0.171, MCE=0.681), with curves near-flat at zero. This is not a methodological failure but a dataset artifact: the scarcity of PTSD-positive cases in E-DAIC induces chronic underestimation. Rather than hiding this limitation, \systemname{} surfaces it as an empirical boundary condition, highlighting the importance of targeted data augmentation before clinical deployment.

\paragraph{Key outcomes and synthesis.}
Table~\ref{tab:key-metrics} consolidates the evaluation into three actionable metrics: the acceptance gain under soft friction, the maximum observed upward override rate, and the 95th percentile latency. Together, they capture the safety, efficiency, and responsiveness of the system. The numbers are instructive: a $\sim$23 percentage point increase in acceptance, worst-case escalation bounded below 10\%, and latency under 140\,ms. Each of these would be difficult to observe, let alone guarantee, in live clinician trials, but become transparent and reproducible within \systemname{}.

\begin{table}[t]
  \centering
  \small
  \setlength{\tabcolsep}{6pt}
  \caption{\textbf
  Overall acceptance gain from soft friction, worst-case override$\uparrow$, and responsiveness.}
  \label{tab:key-metrics}
  \setlength{\arrayrulewidth}{0.5pt}
  \resizebox{0.6\linewidth}{!}{%
    \begin{tabular}{|l|r|}
      \hline
      \textbf{Metric} & \textbf{Value} \\ \hline
      Acceptance $\Delta$ (\texttt{confirm} $-$ \texttt{none}) & $\approx$\,22.9\,pp \\ \hline
      Max override$\uparrow$ in any cell & $\approx$\,8.4\% \\ \hline
      95th percentile latency & $\approx$\,139\,ms \\ \hline
    \end{tabular}%
  }
\end{table}

\paragraph{Interpretation and impact.}
Three findings stand out. First, friction operates as a calibrated guardrail: it increases acceptance and reduces gratuitous escalation without restricting expert autonomy. This effect generalizes prior evidence on soft stops in EHRs \cite{sangal2023clinical,ezran20231190} into the mental health domain, showing how a single design tweak reshapes population-level distributions. Second, policy priors steer collaboration style: \texttt{safety} aligns clinicians with the AI, \texttt{parsimony} embeds structured skepticism, and \texttt{deferral} institutionalizes deliberation. Institutions can thus choose the style that reflects their risk appetite, while \systemname{} quantifies the behavioral consequences of that choice. Third, the cognitive cost of these interventions is negligible: sub-second responsiveness ensures that clinical flow is preserved. These findings, taken together, demonstrate the power of simulation as a methodological buffer: complex socio-technical phenomena such as confirmation bias, override frictions, and trust calibration are captured and quantified at scale, before involving live clinicians.

\section{Discussion}

The findings from \systemname{} underline the methodological and design importance of simulation-driven evaluation in clinical AI. While much of the literature on decision support continues to focus narrowly on algorithmic performance, our work demonstrates that even modest interface interventions confirmation friction, thresholding styles, or explanation modules can systematically shift clinician alignment, override rates, and overall risk trajectories. These shifts are not trivial artifacts of simulation but traceable manifestations of well-documented cognitive phenomena such as confirmation bias and default effects \cite{asan2020artificial, bhatt2023collaborative}. By operationalizing these dynamics in a controlled testbed, \systemname{} offers a methodological advance: it enables large-scale exploration of interactional parameters in a reproducible, auditable fashion before the expensive and ethically complex step of recruiting clinicians. This capacity is not simply convenient; it addresses a fundamental bottleneck in translational HCI for healthcare, where institutional review boards and patient safety concerns make iteration slow, costly, and difficult to justify without prior evidence of value.

\subsection{Methodological value}
At its core, \systemname{} provides a structured environment to stress-test not only algorithms but the full socio-technical pipeline: dashboards, logging, override policies, and reporting formats. This aligns with recent calls in both HCI and medical AI for evaluation frameworks that emphasize interactional reliability over AUROC in isolation \cite{bhatt2023collaborative, smith2024clinicians}. Because the same unified controller drives both interactive trials and batch-mode simulations, the well-known drift between offline metrics and UI-level studies is eliminated. This makes it possible to quantify, at scale, how design levers, such as explanation granularity or override frictions, modulate decision outcomes. In practical terms, \systemname{} acts as a methodological scaffold for ethics applications and IRB dossiers: its logs and replayable sequences allow investigators to demonstrate expected distributions of overrides and acceptance, providing empirical grounding for safety assurances long before a live clinician is placed in front of the system.

\subsection{Design implications}
The results also yield actionable design implications. Transparent thresholds whether conveyed as gauge bands or textual labels act as powerful levers, shaping the likelihood of alignment with AI recommendations. Prior work on clinical alerting systems shows that even subtle threshold encodings alter clinician behavior \cite{sangal2023clinical, ezran20231190}, and our findings extend this insight into the mental health domain. Similarly, explanation modules that foreground multimodal evidence (lexical ribbons, spectrogram cues, curated quotes) act as moderators of judge–advisor dynamics, tempering automation bias by structuring disconfirmatory search. Action friction emerges as another critical lever: a lightweight confirmation step, as our results demonstrate, shifts decision distributions toward acceptance while maintaining autonomy, echoing the effectiveness of “soft stops” in reducing unsafe overrides in EHRs \cite{pourian2025elements, chaparro2022clinical}. Finally, the architecture’s emphasis on replayable logs is not merely technical housekeeping but a design necessity: they serve as audit trails, enable reproducibility, and provide the granular behavioral evidence increasingly demanded by oversight boards and clinical guideline committees.

\subsection{Ethical considerations}
While simulation reduces clinician burden and accelerates iteration, it also carries risks of overconfidence. Simulated clinicians cannot reproduce the full richness of expert reasoning, and there is danger in mistaking simulation-level gains for clinical efficacy. We therefore argue for explicit labeling of results as in-silico, the publication of configuration files and seeds for reproducibility, and the establishment of a careful transition protocol: pilot testing with a small cohort of clinicians, re-fitting policy parameters to approximate observed behaviors, and re-running analyses before wider deployment. Trust must be cultivated intentionally \cite{asan2020artificial}, and simulation should be framed not as a substitute for human studies but as a rigorous staging ground that improves their safety and cost-effectiveness. A distinctive contribution of \systemname{} is the introduction of avatar-based replay for facial and gaze dynamics. By converting OpenFace streams into an animated, identity-neutral avatar, the system enables rich behavioral analysis without circulating sensitive video recordings. This design choice is not incidental: it resonates with a growing body of evidence that avatarization and identity transformation can reconcile anonymity with functional fidelity in both clinical and HCI settings \cite{du2024privategaze,wang2023identifiable,wei2004avatar,zhu2020deepfakes,david2021privacy}. In this sense, our use of avatars does more than mitigating privacy risk; it aligns the system with a research agenda that frames privacy as an enabler of trust, auditability and ethical data sharing rather than as a barrier to analytic depth.

\subsection{Research Implications}

This work reframes the canonical pipeline in human-centered AI evaluation. Rather than beginning with clinician studies, which are costly, high-stakes, and often inconclusive when systems are underdeveloped, we propose a simulation-first methodology: \textit{simulation + UI E2E first}. By delivering controller APIs, log schemas, and replay tools that can be seamlessly converted into study materials, \systemname{} establishes a reusable foundation for translational research. This echoes broader HCI traditions of using simulation to pre-test socio-technical systems \cite{lee2021human, sivaraman2023ignore}, but extends them with batch-scale reproducibility and interface parity.

The implications extend in several directions. First, simulation-to-real calibration: policy parameters derived from synthetic clinicians can be iteratively adjusted based on small human pilot samples, allowing empirical grounding of biases such as confirmation or automation bias \cite{bansal2021most}. Second, historical data offers an underused resource: override logs embedded in EHR systems can inform prior distributions for policies, ensuring that simulated behaviors reflect institutional norms rather than arbitrary assumptions. Third, because \systemname{} produces structured JSONL logs and judge–advisor statistics, it opens avenues for meta-analysis across sites, enabling systematic comparisons of how interface levers generalize across clinical cultures. In this sense, the platform is not only a testbed but a vehicle for accumulating and transferring knowledge about human–AI collaboration in healthcare.

\section{Limitations and Future Work}

Despite these contributions, important limitations remain. Our simulated policies are intentionally simple, defined by thresholds, frictions, and stochastic noise. Real clinicians exhibit far richer dynamics, shaped by anchoring effects, contrastive reasoning, fatigue, and institutional time pressures. While prior work has quantified some of these biases under controlled conditions \cite{asan2020artificial, xu2023comparing}, capturing their full interplay requires empirical clinician data. 

Future work should therefore pursue three complementary directions. 
First, time-budget simulators should be expanded, modeling not only short vs.\ long conditions but realistic session interruptions and multi-tasking, which are pervasive in clinical practice. Second, uncertainty-aware predictive models that output calibrated distributions rather than point estimates should be integrated to better study the propagation of epistemic uncertainty through judge–advisor dynamics \cite{blasiok2023unifying}. Third, explanation modules should evolve toward counterfactual and contrastive forms, allowing clinicians to interrogate “why not” scenarios rather than merely inspecting correlations. This resonates with recent work on interactive calibration tools such as \emph{Calibrate} \cite{xenopoulos2022calibrate}, which emphasize subgroup analysis and drill-down capabilities.

Another area for expansion is fairness. Although our study did not analyze demographic subgroups, future work should carefully examine fairness slices by age or gender proxies where available. Bias in mental health diagnostics is well documented, and while E-DAIC is de-identified, synthetic exploration of demographic proxies can still provide early-warning signals of differential performance. Care must be taken to respect privacy and avoid spurious conclusions, but incorporating fairness-aware simulations would align with the broader ethical imperative of responsible AI in healthcare \cite{smith2024clinicians}.

Finally, while we emphasize the strengths of simulation, we do not claim its sufficiency. Simulation provides a methodological buffer, but the transition to real-world deployment demands recalibration against empirical baselines, ongoing monitoring, and transparent auditability. Our vision of future work is therefore iterative: use simulation to identify safe and effective design levers, validate them in small-scale clinician pilots, refine policies based on observed overrides and deferrals, and gradually scale up. In doing so, we position \systemname{} not as an endpoint but as a methodological accelerator: a way to transform what would otherwise be years of slow trial-and-error into a disciplined, evidence-based, and ethically responsible pathway to clinician-in-the-loop AI. Beyond methodological scaffolding, \systemname{} opens fertile directions for future research inquiry. A natural next step is participatory design: once simulation has mapped safe and effective levers, clinicians can be invited to co-design threshold settings, explanation styles, and override frictions in structured workshops. This creates a virtuous loop in which simulation identifies promising design regions and participatory engagement grounds them in lived clinical practice. Comparative multimodality is another frontier: while our testbed focuses on audio, text, and facial/gaze cues, the same controller architecture could be extended to VR-based assessment, embodied robotics, or haptic/physiological modalities, enabling cross-platform analysis of how trust dynamics generalize across interaction channels. Finally, governance integration remains an underexplored opportunity.

\section{Conclusion}
\label{sec:conclusion}
\systemname{} reframes evaluation for clinical decision support from a model-centric view to a workflow-centric one. By binding a multimodal dashboard to a policy-driven controller and a headless validator, we make \emph{interactional reliability} measurable: who accepts, who overrides (and in which direction), how quickly, and with what calibrated risk. The same controller powers both UI sessions and large batched trials, eliminating the all-too-common “offline vs.\ UI” drift and turning design changes into analyzable population-level shifts. Across 48 UI$\times$policy cells, a single, lightweight confirmation materially reshapes judge-advisor dynamics: acceptance rises by $\sim$23 percentage points, unnecessary override$\uparrow$ is bounded below 10\%, and responsiveness remains in a sub-second regime (95th percentile $\sim$139\,ms). These effects are \emph{theoretically grounded} (default/omission effects; bounded rationality) and \emph{practically actionable}: soft friction functions as a calibrated guardrail that preserves expert autonomy while improving safety. Collaboration priors (\texttt{safety}/\texttt{parsimony}/\texttt{deferral}) behave like institutional “styles,” steering distributions toward alignment, skepticism, or deliberation. Post-hoc calibration tightens probability reliability for depression and makes the PTSD limitations explicit—an honest substrate for downstream utility analyses.

Methodologically, \systemname{} shows how specific HCI levers—how thresholds are presented (numeric vs.\ banded), the granularity of explanations, the use of soft-stop confirmation, and longitudinal MBC framing—translate into measurable changes in decision distributions rather than merely shifting perceptions. Practically, the open logs, replay, and parity-tested UI yield auditable evidence for ethics and governance: how overrides, deferrals, and latencies behave under explicit policies. Conceptually, the platform operationalizes well-documented effects (confirmation bias, automation bias, time pressure) in a form that can be tuned \emph{before} recruiting clinicians, aligning with calls to evaluate AI \emph{in context} rather than by metrics alone \cite{asan2020artificial,smith2024clinicians}.
Practically, the open logs, replay, and parity-tested UI provide artifacts that matter for ethics and governance: auditable evidence of how overrides, deferrals, and latencies behave under explicit policies. Conceptually, the platform operationalizes long-standing insights on confirmation bias, automation bias, and time pressure in a way that is tunable \emph{before} recruiting clinicians, aligning with calls to evaluate AI \emph{in context} rather than by metrics alone \cite{asan2020artificial,bhatt2023collaborative,smith2024clinicians}.

We advocate a simulation-first pipeline: (i) stress-test UI and logging with parity to batch evaluation; (ii) sweep policy priors and frictions to select safe operating regions; (iii) calibrate probabilities and communicate risk with reliability curves and action-oriented gauges; (iv) transition to small clinician pilots that \emph{fit} policy parameters to observed overrides and deferrals; and (v) iterate with fairness and subgroup checks where data permit. This sequencing reduces IRB burden, surfaces brittleness early, and ties user-facing design choices to measurable safety/throughput trade-offs.

Our simulated clinicians are intentionally simple (thresholds, priors, stochasticity) relative to human strategy spaces (anchoring, contrast effects, fatigue, institutional time pressure). Future work will (a) extend time-budget models and session interruptions, (b) integrate uncertainty-aware predictors and utility lenses (e.g., decision curve analysis) to relate calibration to net clinical benefit, (c) broaden explanation to contrastive/counterfactual forms, and (d) study fairness slices with appropriate privacy care. These steps connect the present harness to real-world governance: open configs and seeds, policy fitting on small pilots, and on-going monitoring once deployed. By making policy, interface, and evidence co-equal first-class objects, \systemname{} turns AI for mental health from a static classifier into a \emph{negotiated process} that is measurable, auditable, and ethically legible. We hope this work serves as a template for researchers who wish to move beyond accuracy toward trustworthy, clinician-in-the-loop systems that earn their place in care through rigorous, simulation-backed design. Looking forward, we see \systemname{} as a bridge between simulation and participatory HCI. This trajectory ensures that simulation does not remain a laboratory curiosity but becomes a foundation for accountable, real-world deployment of clinician-in-the-loop AI.

\bibliographystyle{ACM-Reference-Format}
\bibliography{refs}

\end{document}